\documentclass[11pt]{article}
\usepackage{amsmath}
\usepackage{booktabs}
\usepackage{float}
\usepackage{fullpage}
\usepackage{graphicx}
\usepackage{verbatim}
\usepackage{url}
\usepackage{caption}
\usepackage{subcaption}
\usepackage{xcolor}

\graphicspath{{img/}}

\title{Estimating the Effects of Urban Green Regions in terms of Diffusion}
\author{Eric K. Tokuda$^{1}$ \and Florence A. S. Shibata$^{2}$ \and Henrique F. de Arruda$^{1}$ \and Guilherme S. Domingues$^{1}$ \and Cesar H. Comin$^{3}$ \and Luciano da F. Costa$^{1}$ \and Roberto M. Cesar-Jr.$^{2}$ }

\date{%
    $^1$ S\~ao Carlos Institute of Physics, University of S\~ao Paulo, S\~ao Carlos, SP, Brazil \\
    $^2$ Institute of Mathematics and Statistics, University of S\~ao Paulo, S\~ao Paulo, SP, Brazil \\
    $^3$ Department of Computer Science, Federal University of S\~ao Carlos, S\~ao Carlos, SP, Brazil \\
}

\begin{document}
\maketitle

\begin{abstract}
The interaction between cities and their respective green regions
corresponds to an interesting issue that has received growing attention
over the last decades.  These relationships have multiple natures,
ranging from providing habitat for animal life to temperature and
humidity dynamics.  Several methods based on area, size, shape, and distance have
been considered in the literature.  Given that several important
contributions of green regions to urban areas involve temperature,
humidity and gases (e.g.~oxygen) exchanges, which are intrinsically related
to physical diffusion, it becomes particularly interesting to simulate
the diffusion of green effects over urban areas as a means of better
understanding the respective influences.    The present work reports
a related approach.  Once the green regions of a given city are automatically identified
by semantic segmentation and have eventual artifacts eliminated,
successive convolutions are applied as a means to obtain the unfolding
of the diffusion of the green effects along time.  As illustrated, the diffusion 
dynamics is intrinsically interesting because it can be strongly affected 
by the spatial distribution of the green mass.  In particular, we
observed that smaller green regions could substantially contribute to the
diffusion.  The reported approach has been illustrated with respect to the
Brazilian city of Ribeirão Preto, whose small and medium-sized green regions 
were found to complement in an effective manner the diffusion of the green
effects as inferred from the performed simulations under specific parameter
settings.
\end{abstract}

\section{\label{sec:int}Introduction}

As we progress into the 21st century, important issues remain as a subject of scientific
research, including the study of the quality of life in cities and towns, which is not
only a multidisciplinary problem, but also involves many characteristics and dimensions
including but not being limited to mobility, security, and access to common resources such as
parks, hospitals, etc~\cite{farahani2010multiple}.  Among these several aspects, the presence of green areas, their
distribution, access, effects, etc., have constituted the subject of continuous studies
and discussions~\cite{rusche2019mapping,cetin2015using,ovaskainen2002long,fahrig2020several}.  Green areas are important in an urban environment for several reasons,
including: (i) preservation of living species; (ii) contribution to air quality by 
the production of oxygen and filtering of unwanted molecules; (iii) landscape design;
and (iv) leisure.   Therefore, it becomes important to devise means for quantifying
these effects and interactions in an objective manner, so that different city organizations
can be respectively studied and characterized.  

Possible quantitative approaches to studying the effect of green regions in urban areas include the consideration of distance from urban spots to the nearest green regions~\cite{rusche2019mapping}, the distribution of the area~\cite{cetin2015using}, size~\cite{gioia2014size}, shape~\cite{zhang2009estimation}, and distance between green regions~\cite{saura2007new}. Those studies are intrinsically related to the single large or several small (SLOSS) problem~\cite{ovaskainen2002long,fahrig2020several}. Though each of these methods are interesting
and important in themselves, each of them focuses on a relatively specific property of
the studied problem, so that a more complete characterization, modeling and 
understanding of the presence of green areas in cities should benefit from additional,
complementary approaches.

The main purpose of the present work is to develop a framework for quantifying the
potential effects of green regions in urban areas by taking into account the 
physical phenomenon of \emph{diffusion}.  Indeed, several fundamental dynamics in the physical world are directly related and influenced by
diffusion, including thermal propagation, dissemination of gases such as oxygen and
carbon dioxide, as well as the diffusion of elements in water and other liquids.
At the same time, several non-linear dynamics often incorporate a diffusive component.
Therefore, it becomes interesting to consider the effect of green areas into an
urban environment in terms of numerically simulated diffusion of gases, temperature~\cite{maimaitiyiming2014effects},
humidity, as well as other elements to and from the green areas.  This constitutes the main objective of the present work.

The proposed methodology is applied as follows. Satellite images are systematically collected from a city of interest. A sample of the images are manually annotated regarding the green regions and a deep learning-based model is trained to automatically identify green regions in the images. Eventual artifacts generated in the segmentation are removed in a post-processing step. The resulting image can then be used as input for the simulation of the diffusion of green effects, which we henceforth call \emph{green diffusion}. The green regions are used as sources and the diffusion is calculated by successive convolutions with a gaussian kernel. We simulate the green regions role as sources by performing a diffusion with sources, i.e., at each step, the green areas are replenished to full capacity.

In order to illustrate the potential of the reported methodology, we took into
account a satellite image of Ribeirão Preto city (São Paulo State, Brazil).  Having
identified the respective green regions, we also considered two respective versions
without the small and medium sized green regions, therefore allowing the study of the
effect of those areas into the overall obtained diffusion of green effects.  
Interestingly, it was observed that the incorporation of relatively small green
areas can substantially boost the diffusion of green effects.

This article starts by briefly revising some of the main related works, and proceeds
by describing the adopted dataset and segmentation of the respective images regarding their green regions, as well as pre-processing required for removing some segmentation artifacts.  The numerical simulation of the diffusion of effects from the green regions is then described.  A case example with respect to Ribeirão Preto city is then presented and discussed.

\section{\label{sec:rel} Related work}
Satellite images are employed in various studies and can reveal valuable information about given regions~\cite{tong2018increased,bonthoux2019improving,he2020evidence,dawson2018spatial}. One example consists of a study that analyzed the reforestation in China regarding the amount of green vegetation cover along time~\cite{tong2018increased}. Researchers have also investigated the preferences of people regarding different approaches to implementing green areas in cities~\cite{bonthoux2019improving}. Another important aspect that has been studied is the impact of temperature changes on vegetation~\cite{sukopp2003effects}. Furthermore, the relationship between economic indicators and vegetation was studied with respect to some cities~\cite{he2020evidence,dawson2018spatial,tooke2010geographical}. Interestingly, \cite{he2020evidence} shows that economic growth can positively impact the green area level, but vegetation development does not necessarily imply economic growth.  Researchers also studied the possible relationship between vegetation areas and people's education and income~\cite{nesbitt2019has}. The spatiotemporal changes of these green regions, focusing in a particular ecosystem, such as in mountains, is also the topic of several studies, such as in~\cite{bian2020global}. 

Many distinct aspects, intrinsically related to the interplay between cities' green and urban areas, have been explored~\cite{jin2018responses,gioia2014size,janhall2015review,robitu2006modeling}. Some of these studies reveal characteristics regarding the influence of vegetation in cities, such as the impact on temperature~\cite{gioia2014size} and carbon storage~\cite{ren2011relationship}. The information regarding rural areas surrounding the city has also been considered~\cite{cui2019relationships}. Interestingly, in~\cite{gioia2014size} the effect of the sizes of the green regions patches on urban heat islands is studied. More specifically, this study indicates that the green area extension can be positively correlated to the city's temperature. Moreover, the shape of the green patches can also play an important role in the land surface temperature~\cite{zhang2009estimation}. Although many studies have described the importance and impact of vegetation in cities, the authors of~\cite{xing2019role} suggest that the greenery of urban parks could play a minor role in air purification.

Other studies deal with model simulations that incorporate characteristics of the real systems~\cite{robitu2006modeling, takahashi2004measurement,takebayashi2017influence}. For example, in~\cite{robitu2006modeling} the authors deal with the problem of the influence of vegetation and water on the city microclimate by considering computational fluid dynamics. Other aspects were also considered, which include heat conduction and airflow~\cite{takahashi2004measurement}. Another simulation approach, proposed in~\cite{takebayashi2017influence}, considered the air temperature on the relationship between urban and green areas. More specifically, the author studied real measurements and, from this information, proposed a mathematical simulation for better understanding the diffusion effects on regions surrounding green areas.

\section{\label{sec:met} Material and methods}

\begin{figure}[ht]
	\centering
	\includegraphics[width=0.9\textwidth]{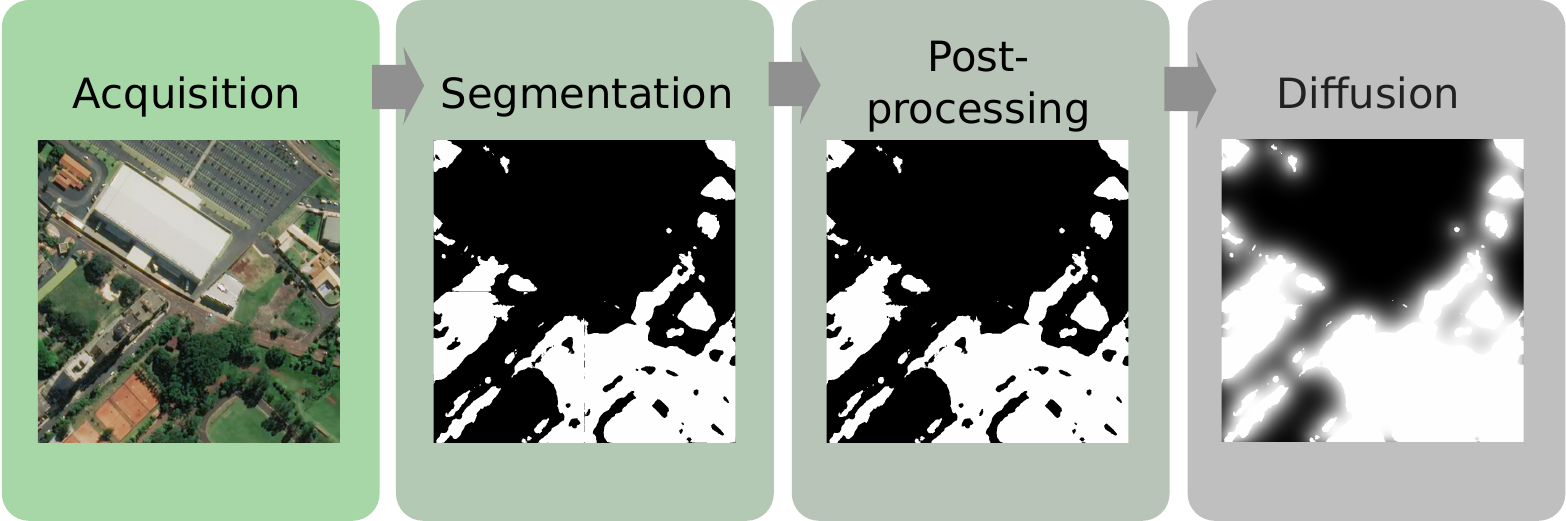}
	\caption{Workflow of the methodology described in the present work.}
	\label{fig:workflow}
\end{figure}

In this section, the diffusion-based methodology for characterizing green areas in satellite images is presented. The main steps of the methodology are: i) dataset acquisition; ii)  segmentation of green areas using a convolutional neural network; iii) post-processing using mathematical morphology; and iv) simulation of a diffusion dynamics using the identified green areas as source. These steps, which are illustrated in Figure~\ref{fig:workflow}, are explained in the Sections~\ref{sec-dataset}, ~\ref{sec-segmentation}, ~\ref{sec-postprocessing} and ~\ref{subsec:methoddiffusion} below.

\subsection{Dataset acquisition}
\label{sec-dataset}

The images used in this study were obtained from Mapbox~\cite{mapbox}, which provides high-resolution images from many different places. The boundary of the region of interest (i.e. urban area of the cities) was manually identified. Due to the tileset convention~\cite{tileset_name}  used in Mapbox API, it is necessary to convert latitudes and longitudes into another set of numerical coordinates before requesting the download of satellite images of the area through the API. In this convention, the entire region is divided into square images with the same sizes (tiles). Each column in this division is mapped to a $Y$ value and each row to a $X$ value. This pair $X,Y$ uniquely identifies a square of the region and covers an interval of coordinates. Providing the list of $X,Y$ coordinates that forms the region, it is possible to retrieve the image of the entire area of interest. If the resolution of the image is high, the coordinate interval captured per tile will be smaller, thus resulting in a larger number of images to cover the total area. The image format and resolution (coverage zoom) can be specified in the API. Specific technical details of the adopted dataset in our experiments are presented in Section~\ref{sec:res}.

A ground truth dataset was manually generated using the online annotator tool Supervisely~\cite{supervisely}. A sample of the images was randomly inspected (preferably at the urban central regions and not at the urban periphery) and polygons covering the class of interest were generated. When polygons are manually annotated in this tool, all points that compose the polygon became available in a json file that is used to generate labels for each class. In our case, using the definition of green area as described in the next section, all of those elements were mapped to the class green, thus generating a binary mask for all annotated images indicating the presence or absence of green for each image.

\subsection{\label{sec:seg} Image segmentation of green areas}
\label{sec-segmentation}

To characterize a satellite image, it is usually necessary to first identify its main relevant structures.  We start this process by green areas segmentation. Many works have approached the problem of segmenting areas of interest in remote sensing images. For instance, \cite{barbieri2011entropy} considered several types of aerial images and, by using the concept of Shannon entropy~\cite{cover1999elements}, sought to distinguish the regions as being urban, rural, or aquatic. More recently, deep learning methodologies have been considered~\cite{amit2016analysis, khan2017forest,ivanovsky2019building,constantin2018accurate,flood2019using, wang2020classification, zhiyu2020greenspace}, such as in~\cite{amit2016analysis}, where a CNN (Convolutional Neural Network) was adopted for automatically identifying natural disasters from satellite images. In another study, by using a CNN architecture and temporal information, changes in forest regions have been identified~\cite{khan2017forest}.

A deep-learning configuration frequently considered in remote sensing studies is the U-Net~\cite{Unet}, which is also based on convolution layers. More information regarding this architecture is presented in Section~\ref{sec:seg}.  Interestingly, though the U-Net was originally considered for dealing with biomedical image segmentation~\cite{Unet}, many studies employed U-Net variations in the analysis of satellite imagery~\cite{ivanovsky2019building, constantin2018accurate,flood2019using,wang2020classification}. Some examples of applications include the detection of buildings~\cite{ivanovsky2019building}, roads~\cite{constantin2018accurate}, and green areas~\cite{flood2019using}.  
 
\begin{figure}[ht]
	\centering
	\begin{subfigure}[b]{0.32\textwidth}
	\includegraphics[width=\textwidth]{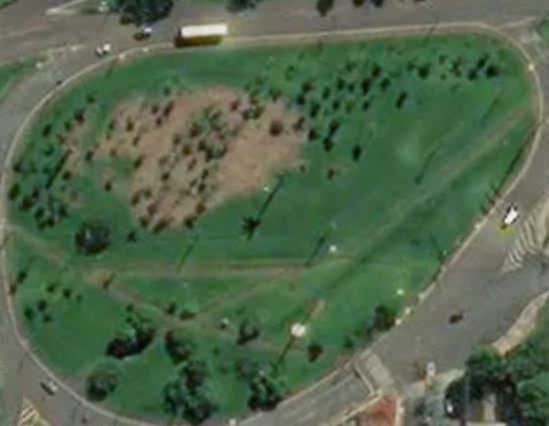}
	\caption{}
 	\label{fig:green_area_ex01}
    \end{subfigure}
    \begin{subfigure}[b]{0.3328\textwidth}
	\includegraphics[width=\textwidth]{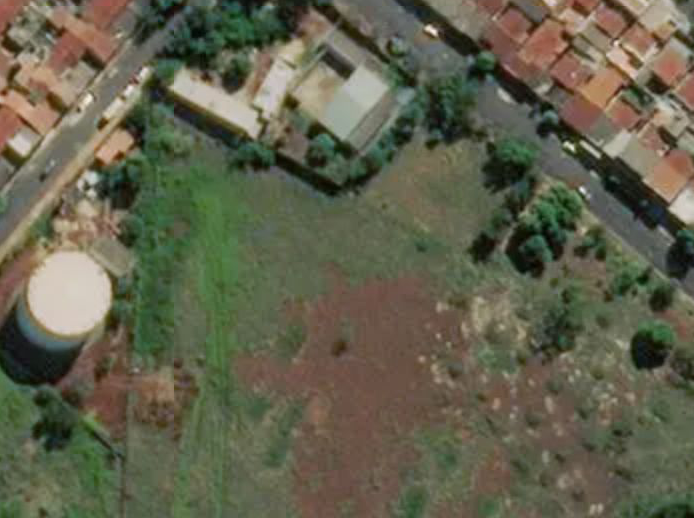}
	    \caption{}
	\label{fig:green_area_ex02}
	\end{subfigure}
	\caption{Example of how ambiguous regions are treated: (a) when 
	trees are present in bare soil, the respective region is considered 
	as a green area; (b) bare soil region devoid of trees is not 
	considered a green area.}
	\label{fig:green_area_example}
\end{figure}

In this study, we focus on inferring the composition of \emph{green areas} in the urban scenery in satellite images, which do not necessarily have a homogeneously green color. Thus, the segmentation of such areas constitutes a challenging task. Here, we consider mostly urban vegetation with green coloration in the definition of Green area: the presence of trees, gardens, squares, yards, flower beds, recreational space with a grass field, trees at roadsides, sidewalks or between residences. In some cases, plantations near the city borders were included. However, there are some areas for which it is difficult to determine the presence or absence of vegetation, such as plantation harvest regions, muddy or barren regions and dry grass with yellow coloration. For these ambiguous regions, we take into consideration the presence of trees as a crucial factor to determine if the area were to be considered as a green area or not. In general rules, the presence of trees will indicate the presence of green. As an example, in the Figure~\ref{fig:green_area_example} there are brown regions surrounded by grass. In Figure~\ref{fig:green_area_ex01}  we can observe the presence of small trees between the region without grass and thus, this region was labeled as a green area. In contrast, the region shown in Figure~\ref{fig:green_area_ex02} also has a brown region but without the presence of any tree and thus was not considered as a green area.

After the construction of the supervised dataset, the data is provided to a supervised learning method to identify the green area regions. The model training process for a segmentation task is challenging because it requires predicting the category of each pixel in the image. Although there are several studies using traditional methods such as Support Vector Machines, powerful deep learning-based methods have been recently developed for this task~\cite{zhu2017deep}. A regular convolutional neural network includes downsampling operations, which may incur a loss of spatial information. The neural network-based approach proposed by Long. et al.~\cite{long2015fully} tries to avoid this problem by transforming the intermediate feature maps back to the size of the input image.  In this study, a U-Net~\cite{Unet} architecture was adopted. The model's architecture is fully convolutional and consists of symmetrical contraction and expansion paths. Additionally, the model associates the contracting path high resolution features with the upsampled output, aiming at preserving spatial location information. By performing pixel-wise classification, the network predicts each pixel belonging to a class by assigning a probability value. Usually, the label assigned to each pixel is that with the highest probability. In our case, we adopted a threshold of 0.6 for considering the pixel as being in the class green.

\subsection{Post-processing using mathematical morphology}
\label{sec-postprocessing}

Because the images are obtained initially as a set of squares, and we employ a tiled-based processing method before assembling the entire image, some artifacts can be respectively implied in the predicted labels. In general, unwanted vertical and horizontal lines result near the square borders. To deal with this problem, we employed a data-based noise reduction post-processing step. First, we apply morphological closing operation using a disk structuring element to reduce noise. Furthermore, morphological closing is also adopted to remove straight line artifacts. More specifically, this step was constrained to the regions composed of pixels that are no further than six pixels from the tile boundaries. In order to remove both the horizontal and vertical unwanted lines, we processed these regions separately, as four rectangles with an overlap at the four corners. These rectangles are named Left (L), Right (R), Up (U), and down (D), according to their positions. With respect to each of the rectangles, kernels with dimensions $6 \times 6$ are defined, as follows
\begin{equation}
    \begin{matrix}
        K_L=
        \begin{bmatrix}
        0 & 0 & 1 & 1 & 1 & 1 \\
        0 & 0 & 1 & 1 & 1 & 1\\
        0 & 0 & 1 & 1 & 1 & 1 \\
        0 & 0 & 1 & 1 & 1 & 1 \\
        0 & 0 & 1 & 1 & 1 & 1 \\
        0 & 0 & 1 & 1 & 1 & 1
        \end{bmatrix}
        ,
        K_R=
        \begin{bmatrix}
        1 & 1 & 1 & 1 & 0 & 0 \\
        1 & 1 & 1 & 1 & 0 & 0\\
        1 & 1 & 1 & 1 & 0 & 0 \\
        1 & 1 & 1 & 1 & 0 & 0 \\
        1 & 1 & 1 & 1 & 0 & 0 \\
        1 & 1 & 1 & 1 & 0 & 0
        \end{bmatrix}
        ,\\ \\
        K_U=
        \begin{bmatrix}
        0 & 0 & 0 & 0 & 0 & 0 \\
        0 & 0 & 0 & 0 & 0 & 0 \\
        1 & 1 & 1 & 1 & 1 & 1 \\
        1 & 1 & 1 & 1 & 1 & 1 \\
        1 & 1 & 1 & 1 & 1 & 1 \\
        1 & 1 & 1 & 1 & 1 & 1
        \end{bmatrix}
        ,
        K_D=
        \begin{bmatrix}
        1 & 1 & 1 & 1 & 1 & 1 \\
        1 & 1 & 1 & 1 & 1 & 1 \\
        1 & 1 & 1 & 1 & 1 & 1 \\
        1 & 1 & 1 & 1 & 1 & 1 \\
        0 & 0 & 0 & 0 & 0 & 0 \\
        0 & 0 & 0 & 0 & 0 & 0
        \end{bmatrix}
        .
    \end{matrix}
\end{equation}

By considering each of these kernels, we compute their convolutions with the respective rectangles. To obtain regions with the same size as the original rectangle, the convolution is performed with zero-padding. The resultant matrix is formed by real numbers. So, in order to convert this matrix to binary numbers, a threshold, $\tau$, is applied, and the values above $\tau$ are converted to 1, otherwise being assigned 0 (here we use $\tau=0.5$). This operation starts from the rectangle L, and the remaining are computed by following the order: R, U, and D. Note that, for the overlapped regions,  this operation is performed twice so as to remove both the horizontal and vertical artifacts.

\subsection{Green diffusion}
\label{subsec:methoddiffusion}

Nature and vegetation are essential in promoting healthy living on earth. Thus, it becomes important to analyze the distribution of the green and non-green regions in its cities. 

Here, we propose a physics-inspired approach for characterizing the interaction between co-existing green and non-green regions, which we call \emph{green diffusion}. A simple process characterizing several natural phenomena is isotropic diffusion, or simply diffusion~\cite{crank1979mathematics}.  For instance, the diffusion of molecules such as oxygen follows this type of dynamics.  This provides one of the main motivations for the present work, namely the simulation of a diffusion process emanating from the green regions into the non-green areas.  If allowed to unfold along time, defining a respective signature, such a diffusion dynamics can provide an indication about the potential influence of the green regions over the other areas.  The specific profile observed in these signatures can vary largely depending not only on the proximity, but also on the \emph{geometry} of the green regions.  Several desired green effects such as humidity, soil preservation, air filtering -- not to mention providing habitat for birds and small animals -- which contribute to environmental and urbanistic aspects, can be reasonably modeled in terms of a linear diffusive dynamics. The green regions are represented as corresponding to sources of diffusion.  More specifically, their values are kept constant at all times.

In a more practical aspect, this method can be computed through successive convolutions with the green region. First, a mask with value one representing the green areas, and zero otherwise, is created by considering the green areas segmented from the satellite image. For each iteration, the image is convolved with a smooth kernel (e.g., Gaussian). This procedure is repeated $n$ times, in order to simulate the diffusion of the green effects throughout the city. 

As an alternative dynamics, we consider the diffusion with sources. This approach starts with the same process employed in the former methodology. However, at the end of each iteration, the intensity of the initially detected green regions is set to one again. Here, the number of iterations is controlled by the following equation
\begin{equation} 
	c = \frac{\text{area}_{\text{green}}}{\text{area}_{\text{total}}},
\end{equation}
where $\text{area}_{\text{green}}$ is the number of pixels reached by the green diffusion, and $\text{area}_{\text{total}}$ is the total number of pixels in the image. The dynamics finish when $t \leq c$, where $t$ is a parameter. 

The methodology can be employed by considering both green and non-green regions as initial sources. In the first case, the dynamics accounts for the effects of vegetation in the city. However, for non-green regions as sources, this methodology can represent cities' growth over areas of vegetation. From a more practical point o view, if we consider the non-green areas as sources of diffusion, the overall amount of green decreases over time. Furthermore, non-green regions, in the initial state, will be kept intact throughout the dynamics, and the original green regions, in turn, will gradually decrease in size. An interesting measure, obtained from these simulations, is the time required to achieve a desired overall green level. 

\begin{figure}[ht!]
    \centering
    \begin{subfigure}[b]{\textwidth}
        \centering
        \includegraphics[width=0.28\textwidth]{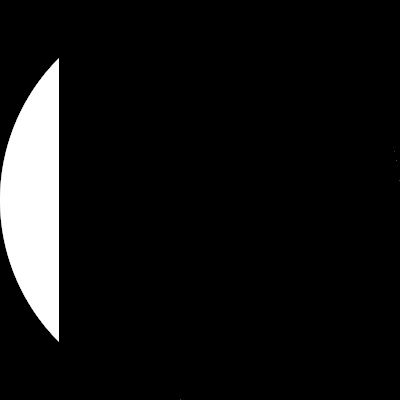}
        \includegraphics[width=0.46\textwidth]{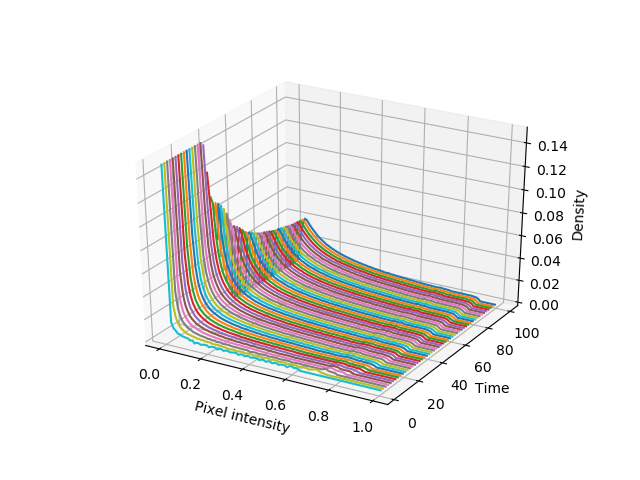} 
        \caption{}
    \end{subfigure}
    \begin{subfigure}[b]{\textwidth}
        \centering
        \includegraphics[width=0.28\textwidth]{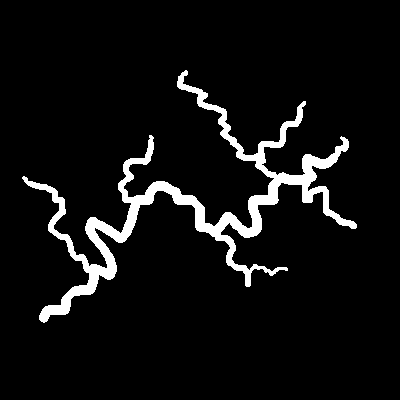}
        \includegraphics[width=0.46\textwidth]{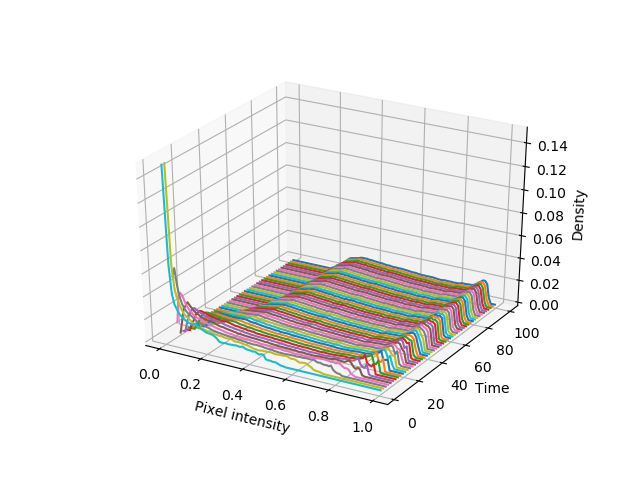}
        \caption{}
    \end{subfigure}
    \begin{subfigure}[b]{\textwidth}
        \centering
        \includegraphics[width=0.28\textwidth]{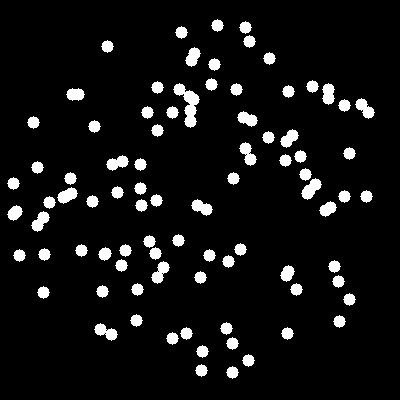}
        \includegraphics[width=0.46\textwidth]{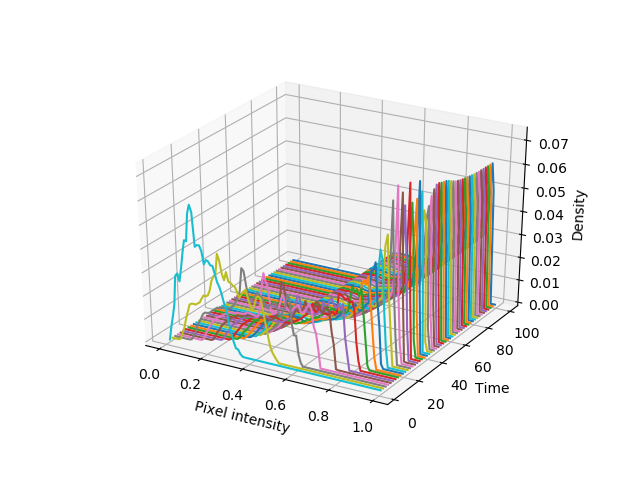}
        \caption{}
    \end{subfigure}
    \caption{Diffusion with sources for three toy models (a), (b) and (c). The first column shows different green area distributions and the second column shows the corresponding histograms across time.  Observe the different
    scales in the histograms density axes, for the sake of better visualization.}
    \label{fig:diffsources}
\end{figure}

Interestingly, though corresponding to a ubiquitous phenomenon in nature, diffusion dynamics 
can vary in surprising manners depending on the spatial distribution and size of the green sources.
This important issue is illustrated in Figure~\ref{fig:diffsources}, which depicts three distinct 
distributions of green regions but with the same total green area.  The three considered
types of green spatial distributions can be summarized as: (a) a single, large connected area, 
for instance as corresponding to parks and squares; (b) a branched distribution of vegetation,
as typically observed around rivers and streams; and (c) a large number of relatively small
vegetation areas, as in gardens and backyards, distributed in a nearly uniform manner. Also shown in Figure~\ref{fig:diffsources} are waterfall diagrams containing the histograms of the diffused green
effect along time, given by the concentration of pixels, respectively to each of the three considered configurations. 

The first important result is that quite distinct green diffusion profiles can be obtained with
respect to each of the three types of spatial distribution.  The case involving a single, large
green area led to histograms indicating large frequency of relatively low diffused effects,
therefore corresponding to a little effective influence of the green region as gauged by 
diffusion.  The branched case yielded a relatively more effective diffusion, with histograms
being shifted to the right.  However, the third type of green spatial distribution, involving
a relatively uniform spreading of small vegetation areas, resulted in the most effective
diffusion of green effects, being substantially distinct from the other two types of distribution
types.  This result is particularly important with respect to the present work because: (i)
it shows that the diffusion can critic and abruptly vary in non-intuitive manner
depending on the size and spatial spread of the green areas; (ii) the most effective diffusion
of a total green area takes place for several small regions distributed in nearly uniform
manner.  

\section{\label{sec:res} Results and discussion}

\subsection{Dataset and implementation details}

To illustrate the proposed methodology, we select a Brazilian city, Ribeirão Preto, from São Paulo state, with 711,825 inhabitants. It was founded in 1856, and has a total area of 650.916$km^2$. Furthermore, the region has a tropical climate. 

Our methodology starts by acquiring satellite images, identifying green areas, and analyzing a dynamical process considering this topology. The satellite images used in this study were obtained using the Mapbox Raster Tiles API~\cite{raster_tile}, with the zoom at level 18, which corresponds to the resolution of approximately 0.6 m/pixel, providing high-resolution images from the entire region of the city. With this resolution, a total of 49911 tile images were acquired to compose the entire region of the city. The three-band RGB color images, with size 512 x 512 each, were obtained in the year 2018.

As mentioned in Section~\ref{sec-dataset} a ground-truth dataset was generated using the online annotator tool Supervisely~\cite{supervisely}. A total of 150 images were annotated and for each image, polygons covering the area of interest were generated and mapped into binary labels. Applying rotations of 90, 180 and 270 degrees to augment available data for model training, a total of 600 images were used and then divided into training, validation and test sets with $70\%, 20\%$ and $10\%$ split proportions respectively. 

The model was implemented in python using Keras and Tensorflow libraries. The Adam optimizer was used in the network training with a learning rate 0.00015 and the dice coefficient loss function. The number of epochs was set to 100 and the mean dice coefficient reached $95\%$  on the test dataset. Here we applied a threshold of 0.6 to consider a pixel as belonging to a green area. The segmentation experiments were carried out in Graphics Processing Units (GPU) GeForce GTX 1080 Ti. Training the neural network took approximately 3 hours in this hardware. 

\begin{figure}[ht!]
	    \centering
    \begin{subfigure}[b]{0.44\textwidth}
	\includegraphics[width=\textwidth]{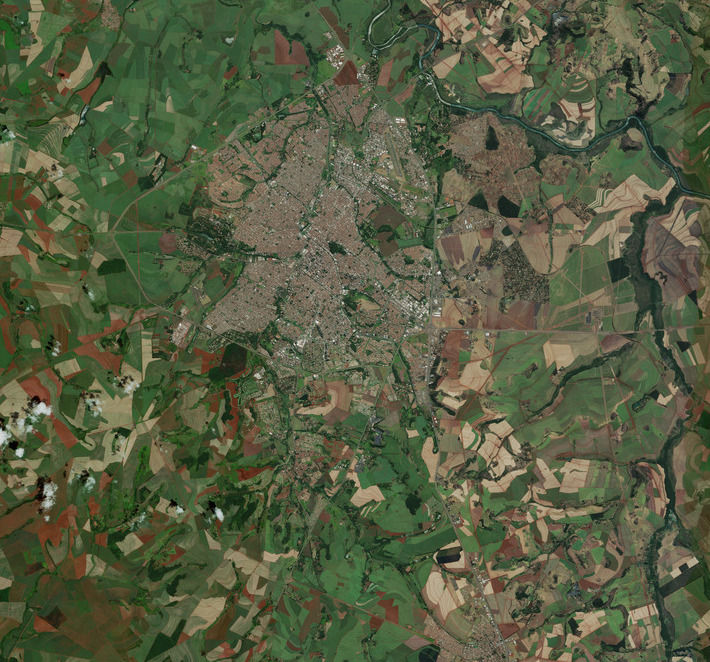}
	\caption{Satellite image}
    \end{subfigure}
    \begin{subfigure}[b]{0.44\textwidth}
	\includegraphics[width=\textwidth]{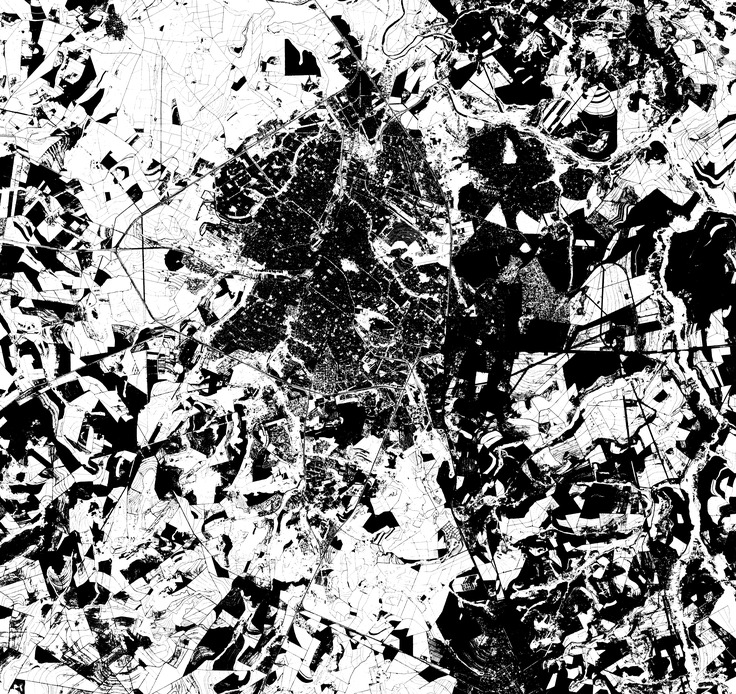}
	    \caption{Green segmentation}
    \end{subfigure}\\
    \begin{subfigure}[b]{0.44\textwidth}
    \includegraphics[width=\textwidth]{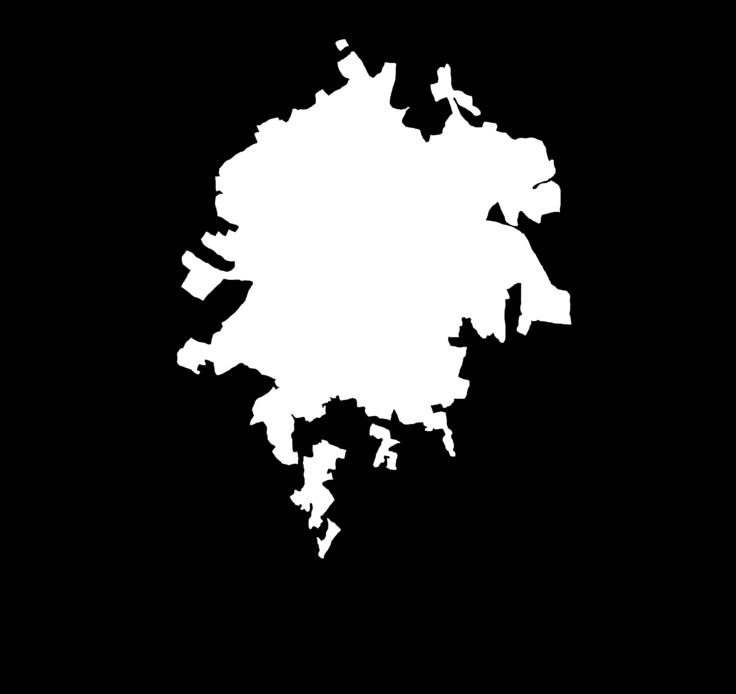}
	\caption{Mask of the urban area.}
    \end{subfigure}
    \begin{subfigure}[b]{0.44\textwidth}
    \includegraphics[width=\textwidth]{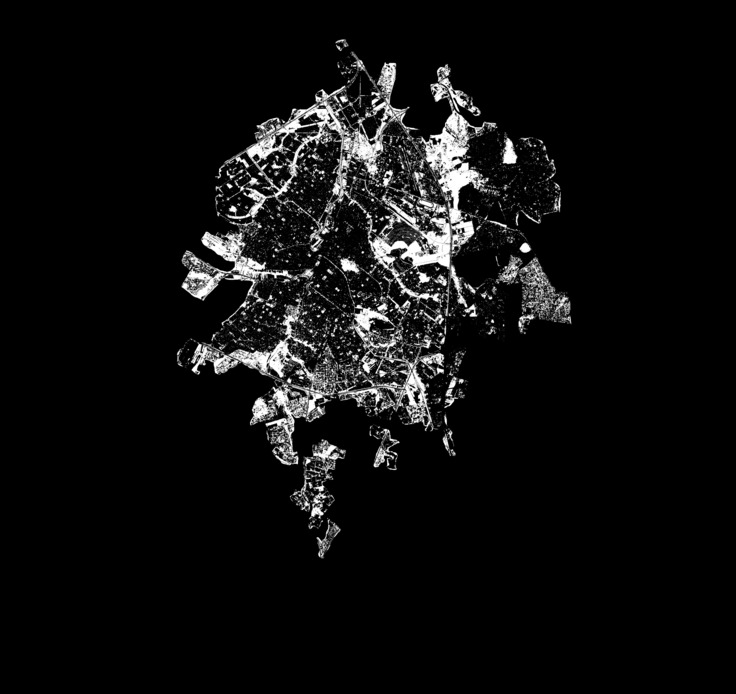}
	    \caption{Urban area.}
    \end{subfigure}\\
    
	\caption{Satellite image (a) and corresponding segmentation of green areas (b) from the city of Ribeirão Preto. Satellite image obtained from Mapbox\textsuperscript{\textcopyright}~\cite{mapbox}. In order to
	remove the rural areas, a mask (c) is applied to the segmentation
	result in (b), yielding (d).}
    \label{fig:segmentationRP}
\end{figure}

An example of segmentation of the green regions in the satellite image
of Ribeirão Preto (Figure~\ref{fig:segmentationRP}a) according to the adopted methodology (Section~\ref{sec-segmentation}) is shown in white in Figure~\ref{fig:segmentationRP}(b). Rural areas were removed using a mask (Figure~\ref{fig:segmentationRP}c) that is applied to the segmentation result, yielding the result shown in Figure~\ref{fig:segmentationRP}(d). This mask was obtained by manual segmentation of the estimated urban region.

\subsection{Diffusion of the green effects}

The process described in Section~\ref{subsec:methoddiffusion} was applied to the considered green areas of the urban region of Ribeirão Preto, comprised by the adopted mask. The image was downsampled by $8 \times$ the original image, resulting in a frame of size $7360 \times 6944$. A gaussian kernel of mean 0 and standard deviation of 201 was considered in the diffusion. The process was evaluated for 100 steps.

\begin{figure}[ht!]
    \centering
    \begin{subfigure}[b]{0.45\textwidth}
    \centering
        \includegraphics[width=\textwidth]{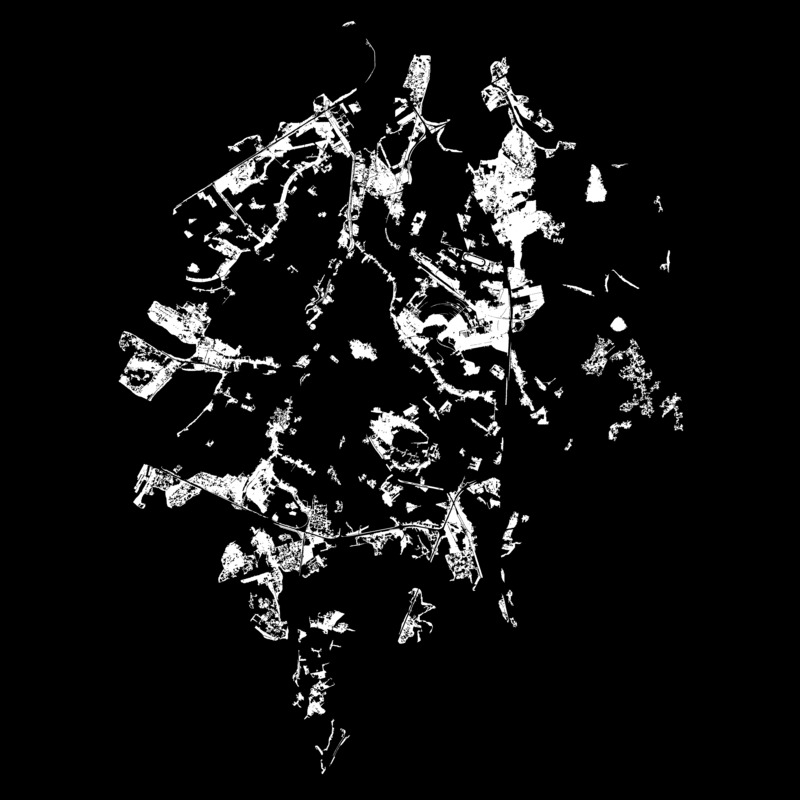}
    \caption{Removal of small green areas.}
    \end{subfigure}
    \begin{subfigure}[b]{0.45\textwidth}
    \centering
        \includegraphics[width=\textwidth]{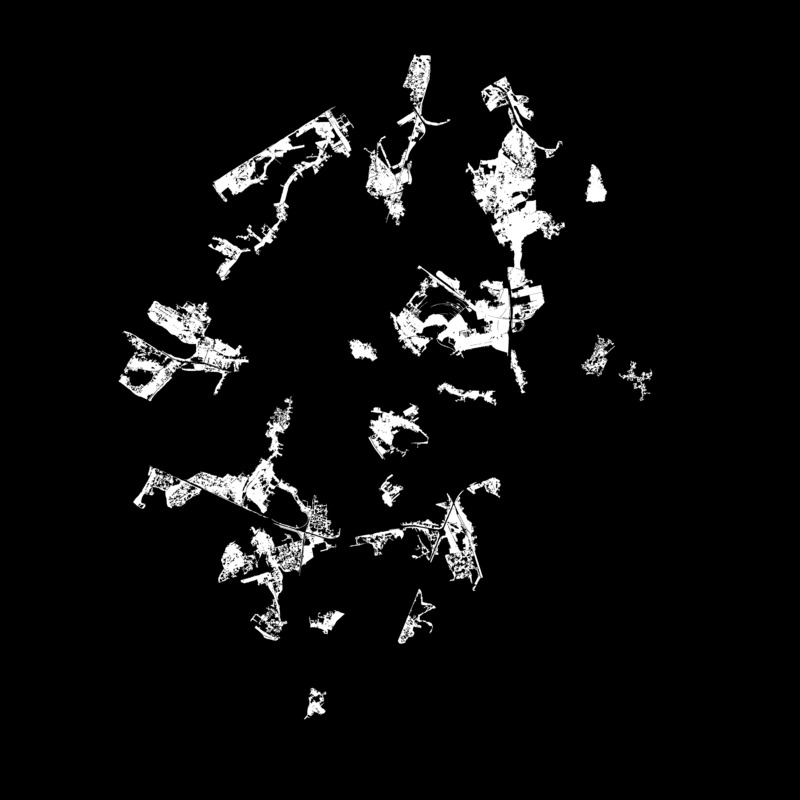}
    \caption{Removal of small and medium green areas.}
    \end{subfigure}
    \caption{Filtering by the area of the connected components, so as to study the influence of green region areas on the overall diffusion. In (a), small areas (smaller than or equal to five times the average size of a tree ($5 \times 100$ pixels) were removed while in (b), small and medium-size areas (smaller than or equal to twice the average size of a block ($2 \times 1000$ pixels) have been removed.}
    \label{fig:filteringarea}
\end{figure}

\begin{figure}[ht!]
    \centering
    \begin{subfigure}[b]{\textwidth}
    \centering
    \includegraphics[width=.33\textwidth]{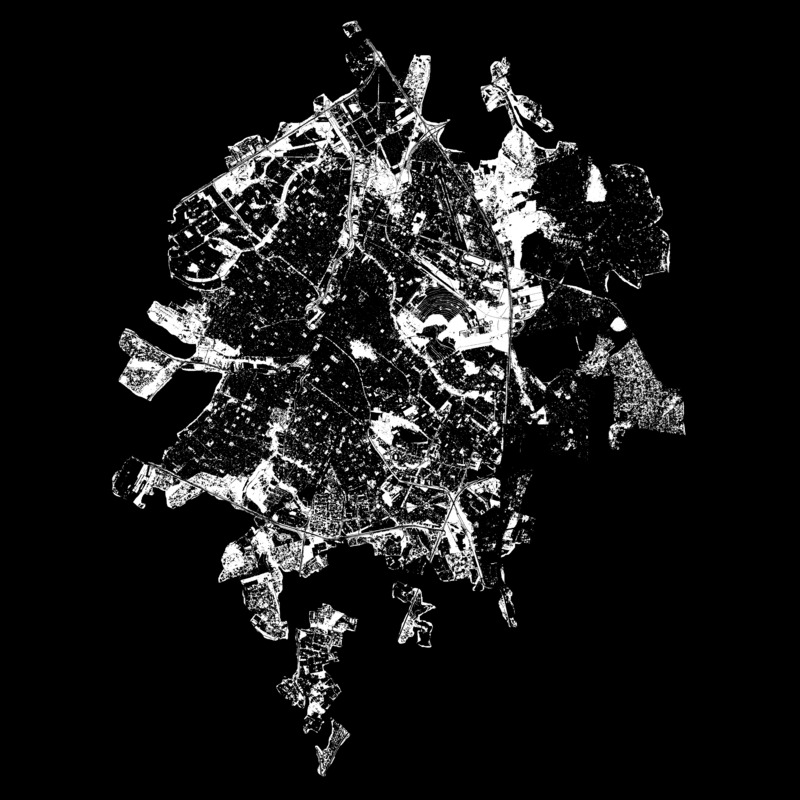}
        \includegraphics[width=.45\textwidth]{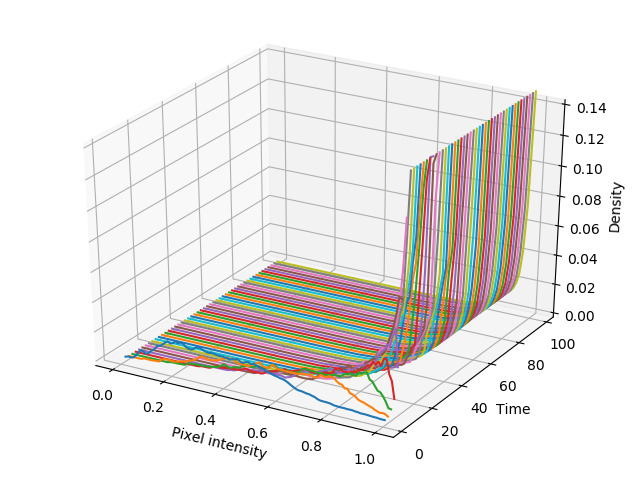}
    \caption{Original map.}
    \end{subfigure}\\
    \begin{subfigure}[b]{\textwidth}
    \centering
        \includegraphics[width=.33\textwidth]{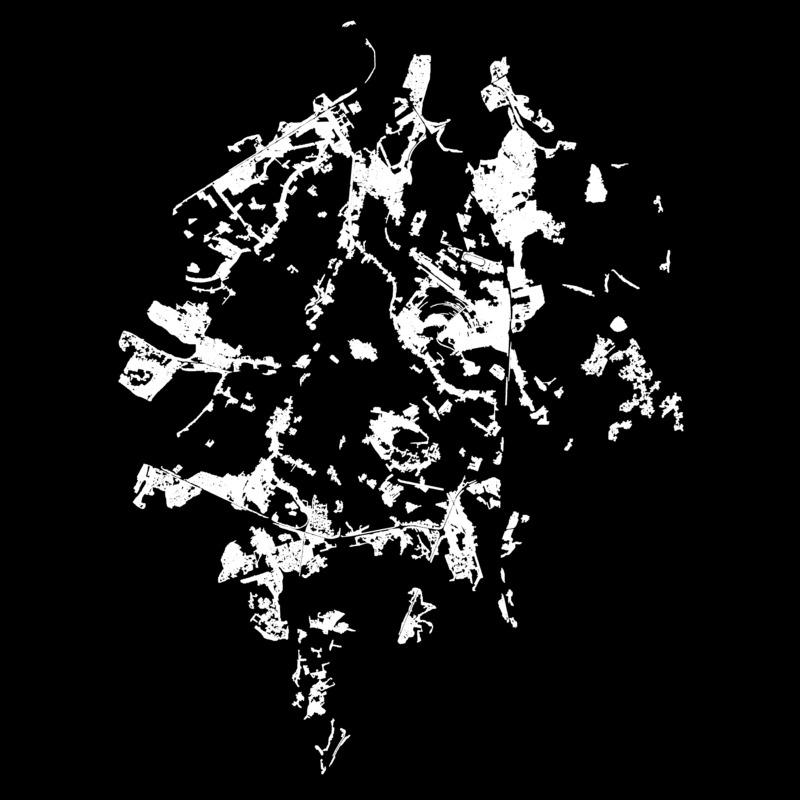}
        \includegraphics[width=.45\textwidth]{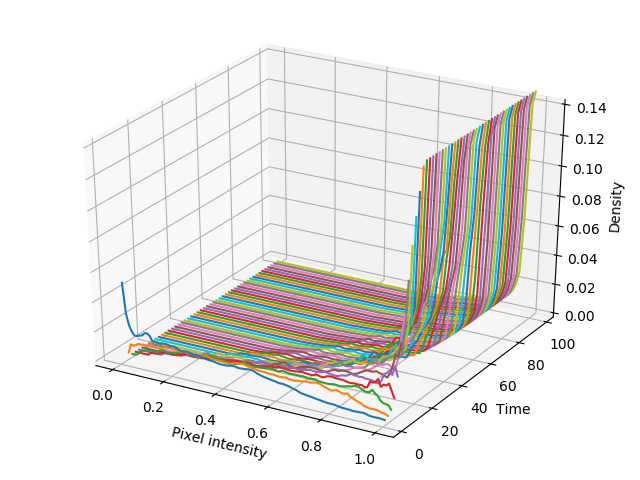}
    \caption{Removal of small green areas.}
    \end{subfigure}\\
    \begin{subfigure}[b]{\textwidth}
    \centering
        \includegraphics[width=.33\textwidth]{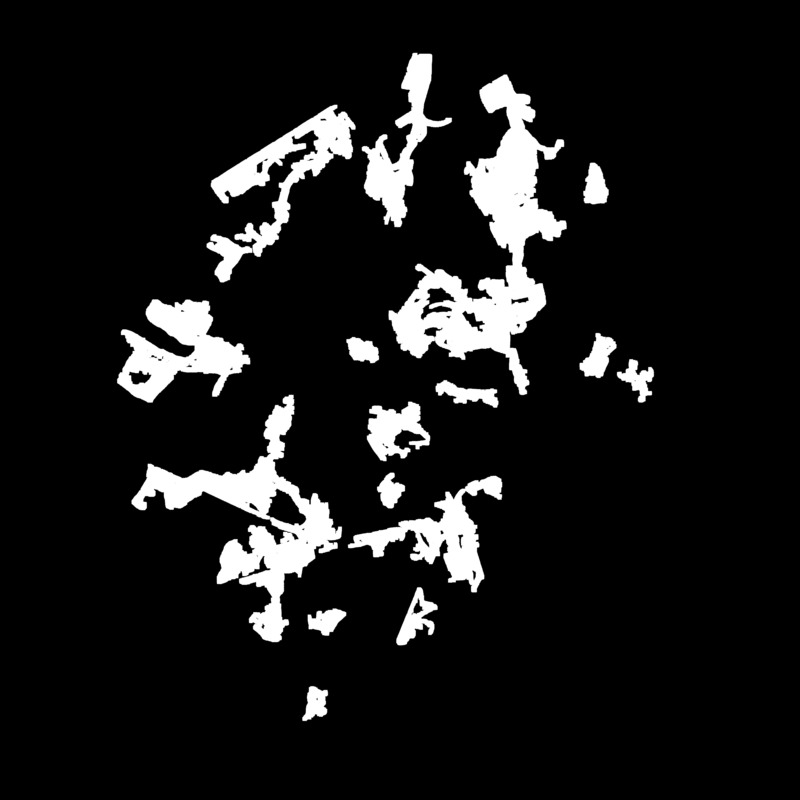}
        \includegraphics[width=.45\textwidth]{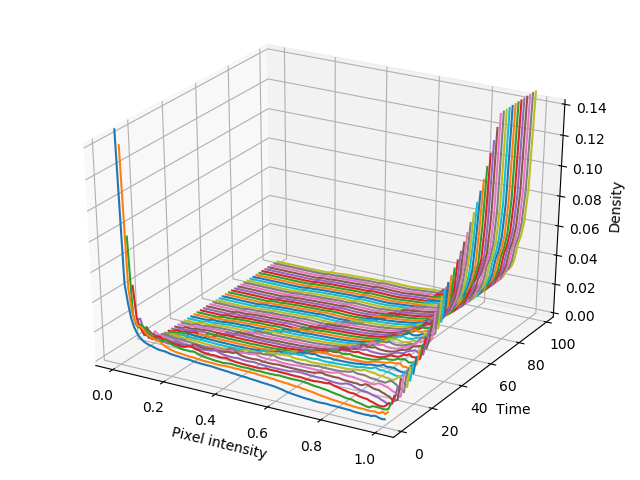}
    \caption{Removal of small and medium size green areas.}
    \end{subfigure}
    \caption{Diffusion of green effects with sources considering the city of Ribeirão Preto. The first column the originally detected green regions (a), as well as the two considered dilated respective versions of the previous image after removal of small (b) and medium (c) size green areas.  The second column contains the corresponding distribution of green along time in the green diffusion.}
    \label{fig:diffRPcuts}
\end{figure}

In order to study the effect of the green region area on the overall diffusion, we consider not only all the identified green area, shown in Figure~\ref{fig:diffRPcuts}(a), but also two additional images obtained by removing the green regions with area smaller than or equal to $5A$, where $A=100$ is the typical size of a tree (Figure~\ref{fig:filteringarea}(a)) and in Figure~\ref{fig:filteringarea}(b) by removing medium-sized green regions with area smaller than or equal to $2B$, where $B=1000$ is the typical size of a block. However, given that the removal of the small and medium sized green regions effectively reduces the overall green area in those images, it would not be proper to compare the diffusion effects between them.  In order to ensure identical overall green areas in the three compared images, we dilated the two thresholded images until they all had the same area as the original image in Figure~\ref{fig:diffRPcuts}(a).  These dilated images are depicted in~\ref{fig:diffRPcuts}(a) and~\ref{fig:diffRPcuts}(b), respectively to the removal of small and medium sized green regions.

Figure~\ref{fig:diffRPcuts} presents the green diffusion results obtained for: (i) the full resolution image of green regions; (ii) the same region after removing connected components smaller than or equal to $5A$ and (iii) the region in (i) after removing all connected component smaller than or equal to $2B$.  The consideration of these three situations allows the verification of the effect of the smaller regions on the overall diffusion dynamics.

The results obtained in all the three considered situations confirm the effect of
diffusion of the green effects, verified in the progressive right-shift of the mass of
the histograms in the waterfall plots along time.  However, in case (i) the diffusion takes place at a much faster rate, reaching saturation after about 20 diffusion steps.  Contrariwise, the situation in (iii) is completely different, with rather slow right-shifting of the
mass in the histograms.  Situation (ii) represents an intermediate result, but with saturation being observed only after 40 diffusion steps.

These results are in full agreement with our previous experiments in Section~\ref{subsec:methoddiffusion}, corroborating further the particularly important effect of the spatial distribution of the green areas on the velocity in which the green effects diffusion can take place.  From a real-world perspective, these results suggest that smaller patches of vegetation may play an important role as far as diffusion of green effects as modeled in the present work are concerned.

\subsection{Validation of the detected green areas}

\begin{figure}[ht!]
	    \centering
    \begin{subfigure}[b]{0.33\textwidth}
	\includegraphics[width=\textwidth]{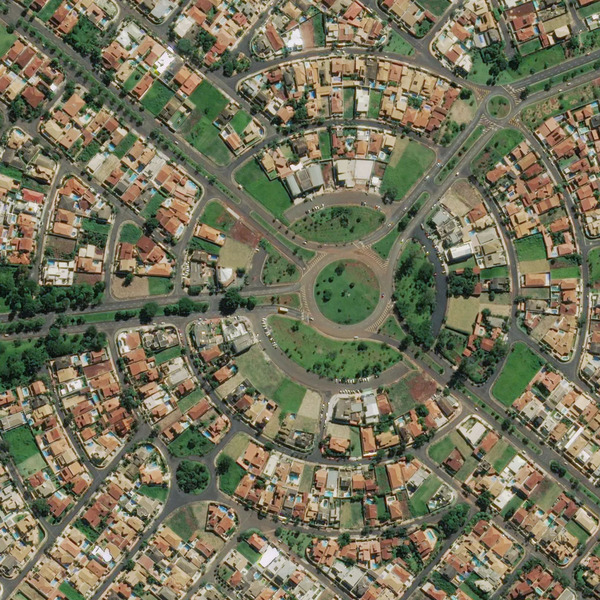}
	\caption{Satellite image of a region of Ribeirão Preto.}
    \end{subfigure}\\
    \begin{subfigure}[b]{\textwidth}
	    \centering
	\includegraphics[width=.33\textwidth]{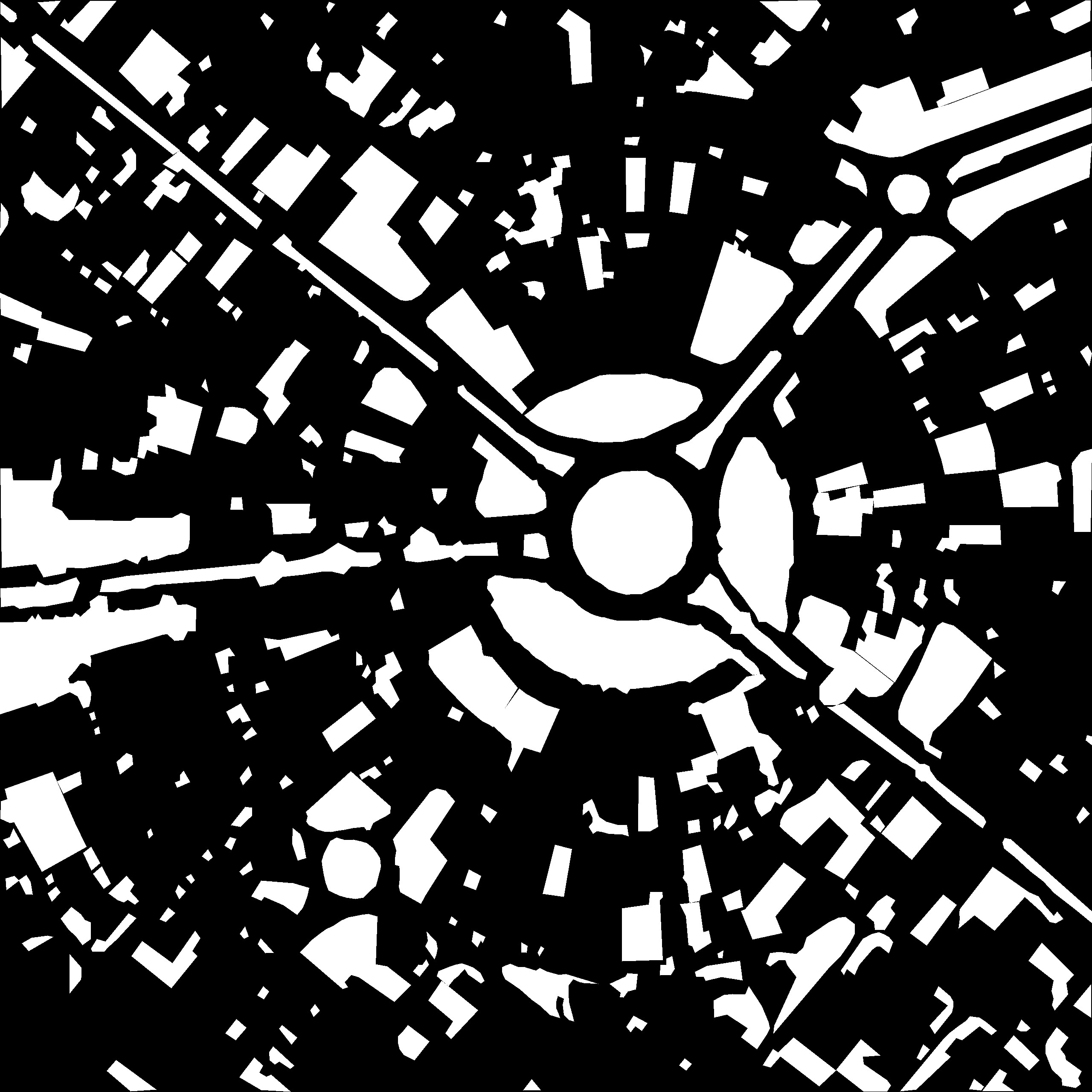}
	\includegraphics[width=.49\textwidth]{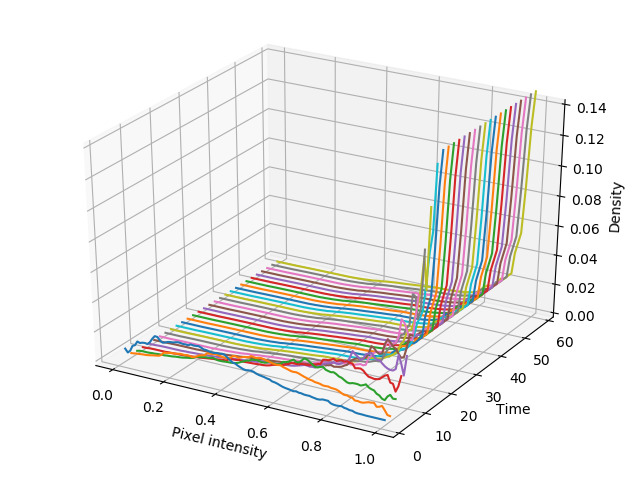}
	\caption{Diffusion considering the manually annotated green areas.}
    \end{subfigure}\\
    \begin{subfigure}[b]{\textwidth}
	    \centering
	\includegraphics[width=.33\textwidth]{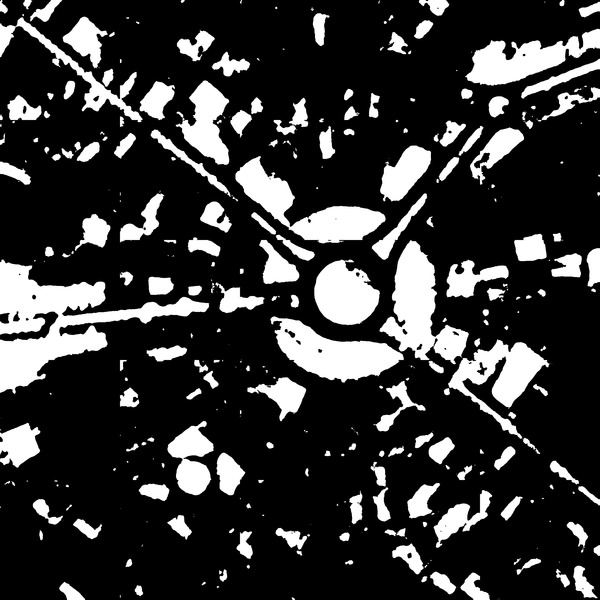}
	\includegraphics[width=.49\textwidth]{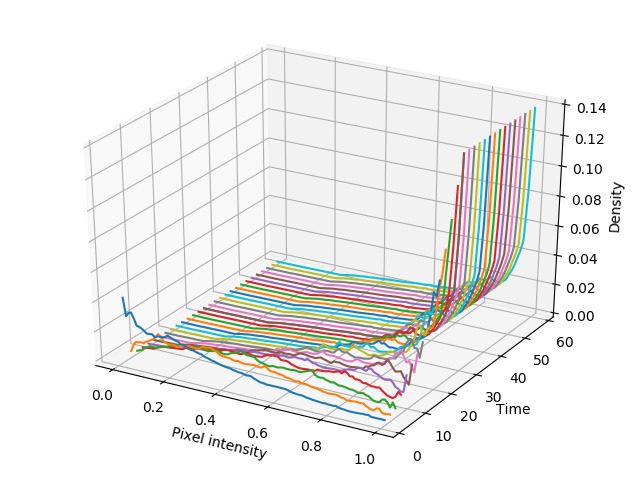}
	\caption{Diffusion considering the green areas automatically obtained.}
    \end{subfigure}
	\caption{Analysis of the quality of the segmentation. }
    \label{fig:validation}
\end{figure}

It is important to analyze the impact of the quality of the segmentation on the results obtained by the diffusion process. To this end, portions of the city with different visual distributions of green have been chosen. We considered regions equivalent to $5 \times 5$ tiles, resulting in images of size $2560 \times 2560$. These regions were manually annotated and we simulated the green diffusion on these images. The results were individually compared to the same process but considering the green areas identified by the DL model. Figure~\ref{fig:validation} shows one of the chosen regions and the corresponding diffusion process on the manually identified green areas and on the automatically identified green areas. It can be seen significant accordance of the two results, which indicate the good quality of the green areas identified by the DL model.

\section{\label{sec:conc} Conclusions}

The presence of green patches in urban regions motivates several important and interesting research possibilities, which may involve temperature, carbon storage, economic value of the properties, etc.   

Given that a good deal of the influence of green regions in an urban environment
take place through diffusion (e.g.~temperature variations, oxygen and other gases
displacement, as well as air humidity), it becomes interesting to simulate the 
diffusion of these effects from green regions in segmented areas of cities.  

The present work reported a related approach.  First, satellite images have their green regions identified and treated as sources of diffusion.   Then, diffusion is simulated by using a numerical approach, more specifically successive convolutions with a
gaussian kernel.  After the diffusion has become nearly stable, histograms 
of the respectively achieved concentrations at each of the considered pixels are obtained. These histograms provide a particularly effective means for gauging the diffusion of green effects given particular regions and cities of interest.

One particularly interesting question implied by diffusion dynamics concerns the
fact that distinct spatial distributions of the green sources can have an unpredictable
and even surprising effect.  This has been illustrated with respect to three hypothetical images containing the same total green area distributed as a single compact area, a branched structure remindful of rivers and streams, as well as an almost uniform distribution of small patches of green regions.  Interestingly, the histograms of green effects obtained in each of these three cases are markedly distinct, with the latter configuration accounting for the most effective diffusion.

As a case example illustrating the proposed methodology, we take into account 
the Brazilian city of Ribeirão Preto.  In order to study the effect of the size 
of the the green regions, we thresholded the respectively obtained green regions 
so as to remove the small and medium sized areas, while special care (dilation of
the threshold images) is applied in order that all the three compared 
configurations have the same total green area.  The respectively obtained 
simulation results suggest that the incorporation of small and medium-sized 
green regions could have a significant effect in boosting the diffusion of 
green effects.

It should be kept in mind that the obtained results are respective to the dynamics
of diffusion as simulated numerically considering specifically adopted city and 
parameter configurations in simplified settings (e.g.~absence of wind and other
external influences).  In addition, it should be kept in mind that 
the considered diffusion dynamics is more closely related only to some of the
various important roles of green regions in urban areas.  In other words,
the obtained results related potentially to the aspects of temperature,
humidity and gases exchanges, not being directly associated to other important
issues such as providing shelter for animal life, leisure, landscaping,
rain absorption, biodiversity, etc.  These diverse dynamics present other
characteristics not modeled by diffusion, such as the need for larger green
areas required for animal life, etc.  Additional research is required for
validating and extending the reported results with respect to more realistic
and diverse configurations.

The methodology and results described in the present work pave the way to several interesting subsequent related studies.  For instance, it would be interesting to
compare the diffusion dynamics with other types of physical dissemination, such as
progressive dilation of regions as observed when expanding or reducing the areas
of parks.  Another interesting possibility regards the incorporation of effects
caused by external effects, such as wind.  It would also be interesting to 
enhance the segmentation methodology so as to be able to distinguish between
different types of green regions (e.g.~trees, lawns, and bushes), as well as
their respective conditions (e.g.~dry or wet seasons).

\section*{Acknowledgments}


Eric K. Tokuda thanks Fapesp for sponsorship (grant \#2019/01077-3). Henrique F. de Arruda acknowledges FAPESP for sponsorship (grant \# 2018/10489-0).  Cesar H. Comin thanks FAPESP (grant no. 18/09125-4) for financial support.  Luciano da F. Costa thanks CNPq (grant \# 307085/2018-0) and NAP-PRP-USP for sponsorship. This work received financial support of the S\~ao Paulo Research Foundation (FAPESP) grant \#2015/22308-2. We are also grateful to Conselho Nacional de Pesquisa (CNPq) and CAPES. 

\bibliography{main}
\bibliographystyle{plain}

\end{document}


\maketitle

\begin{figure}[ht!]
    \centering
        \includegraphics[width=.22\textwidth]{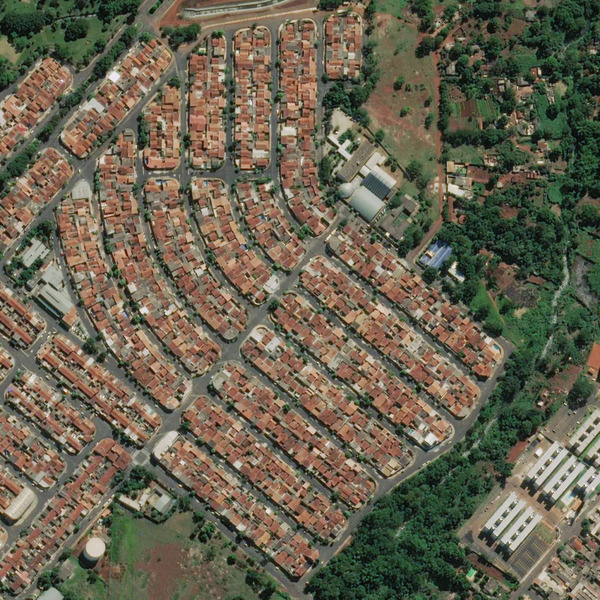}
        \includegraphics[width=.3\textwidth]{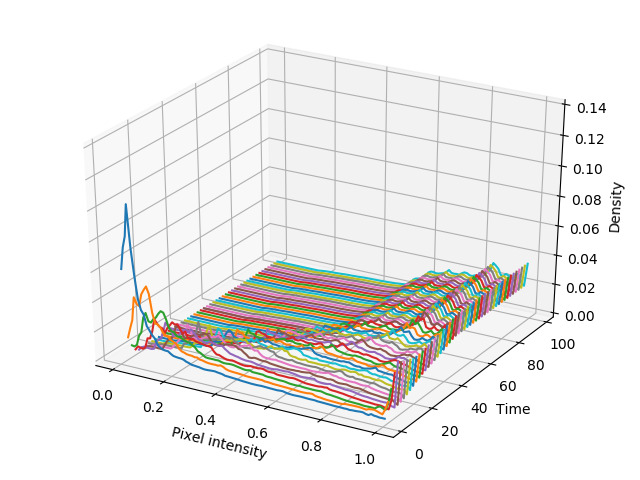}
        \includegraphics[width=.3\textwidth]{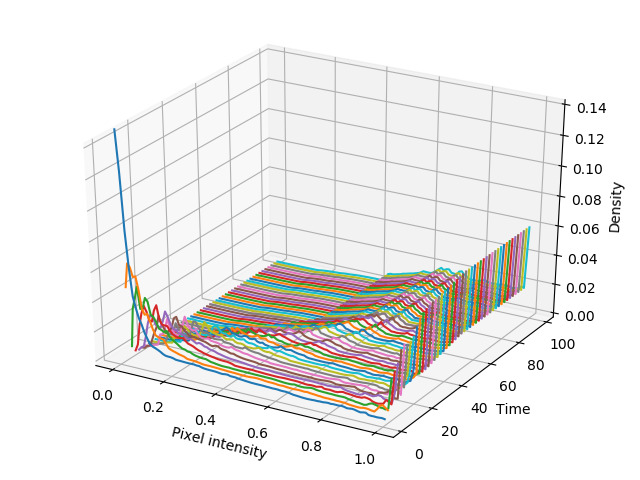} \\
        \includegraphics[width=.22\textwidth]{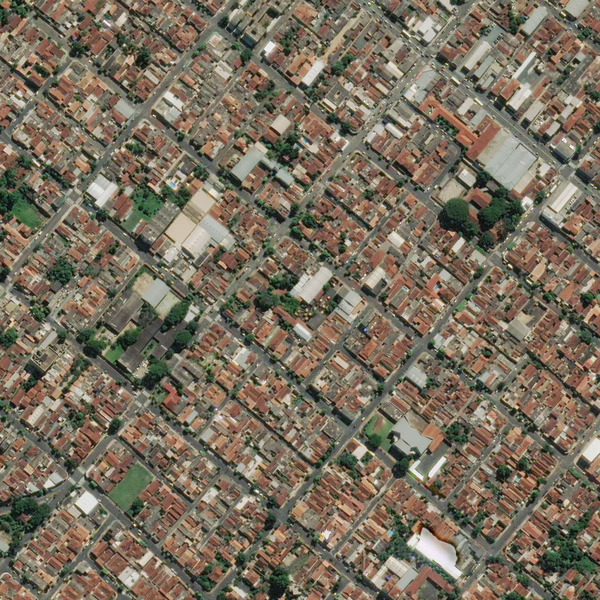}
        \includegraphics[width=.3\textwidth]{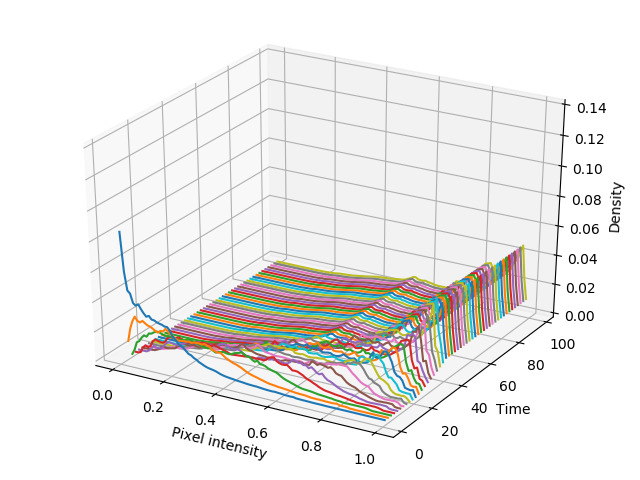}
        \includegraphics[width=.3\textwidth]{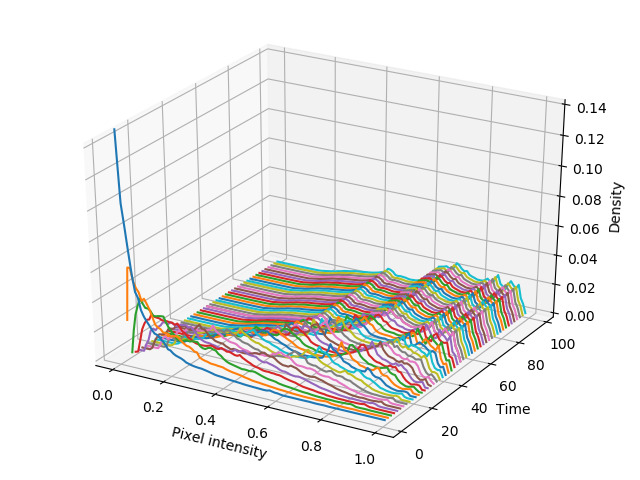} \\
        \includegraphics[width=.22\textwidth]{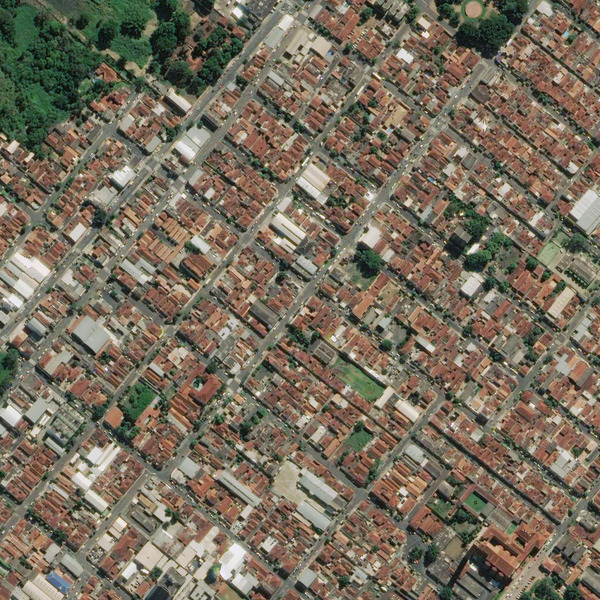}
        \includegraphics[width=.3\textwidth]{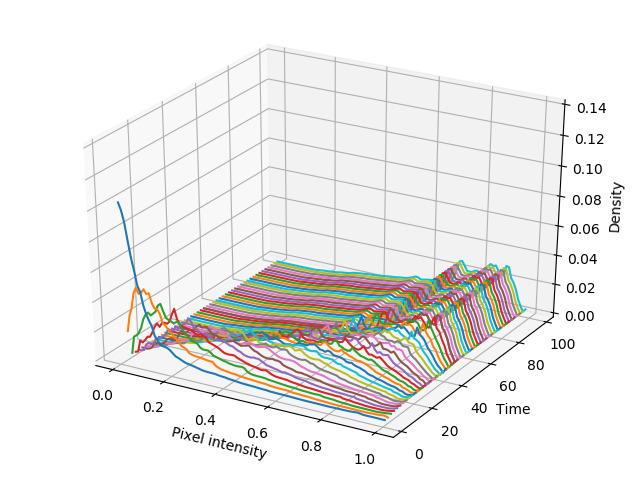}
        \includegraphics[width=.3\textwidth]{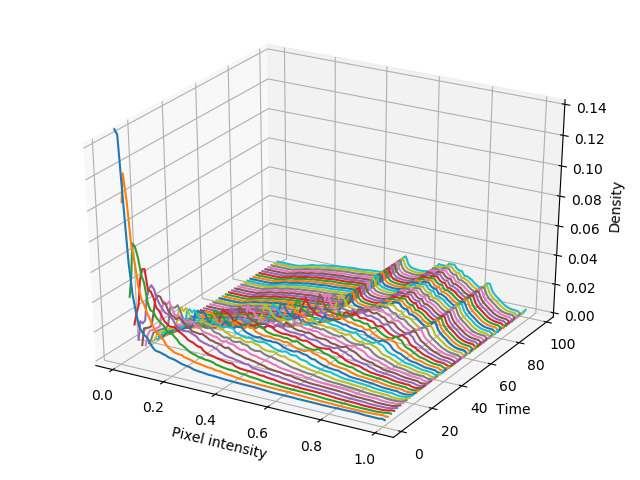} \\
        \includegraphics[width=.22\textwidth]{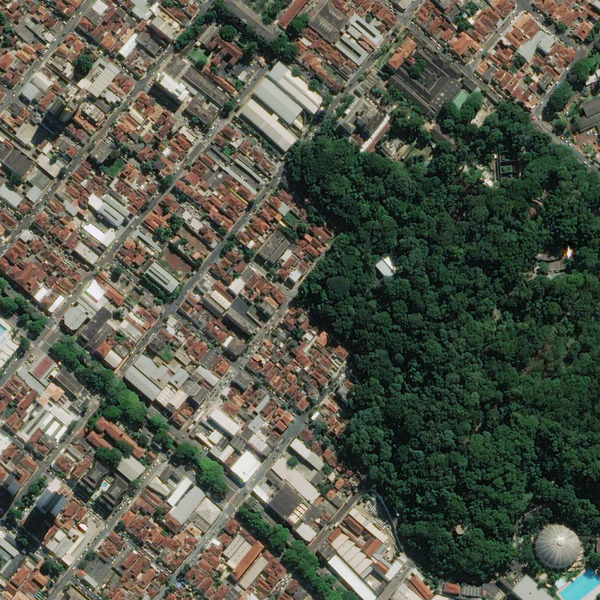}
        \includegraphics[width=.3\textwidth]{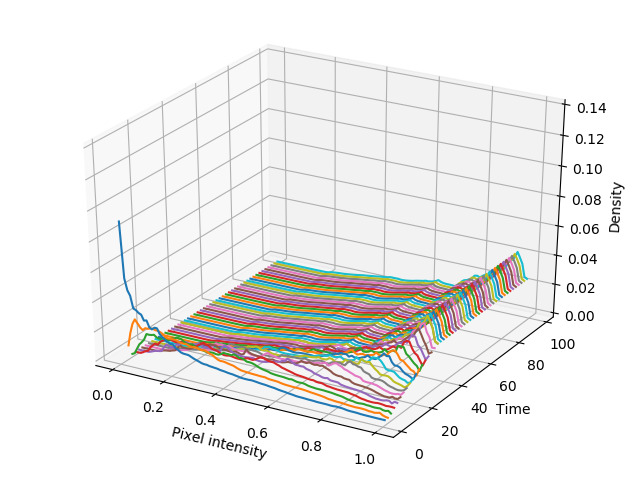}
        \includegraphics[width=.3\textwidth]{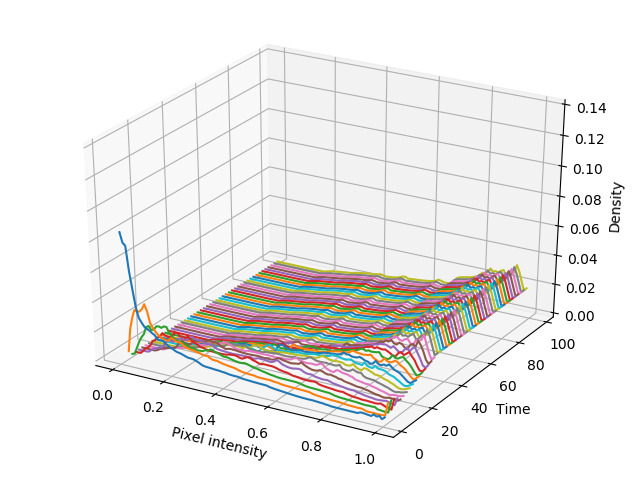} \\
        \includegraphics[width=.22\textwidth]{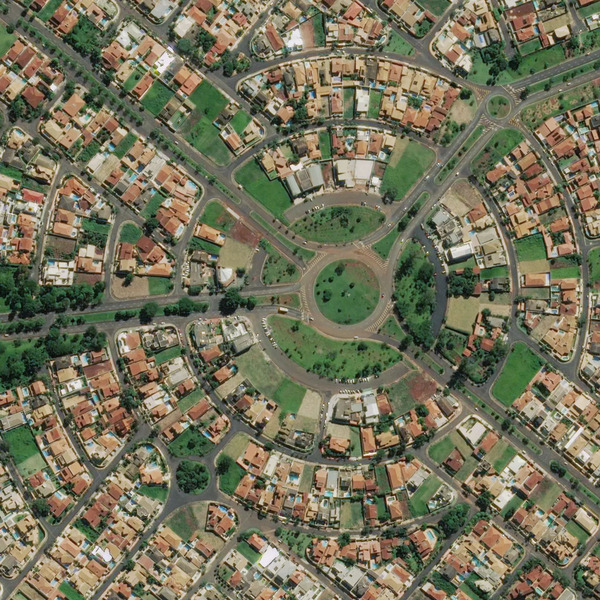}
        \includegraphics[width=.3\textwidth]{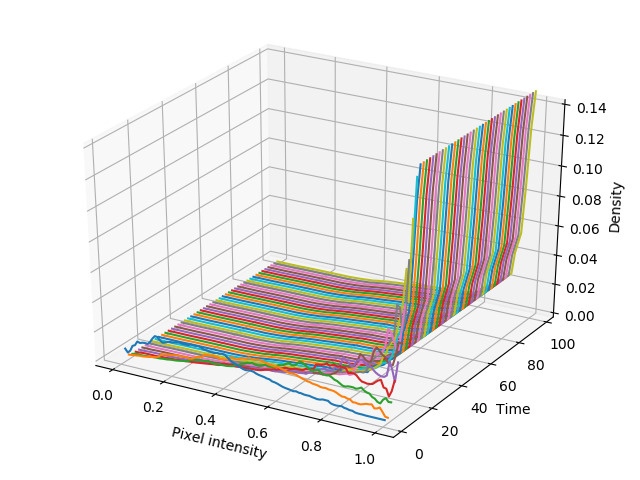}
        \includegraphics[width=.3\textwidth]{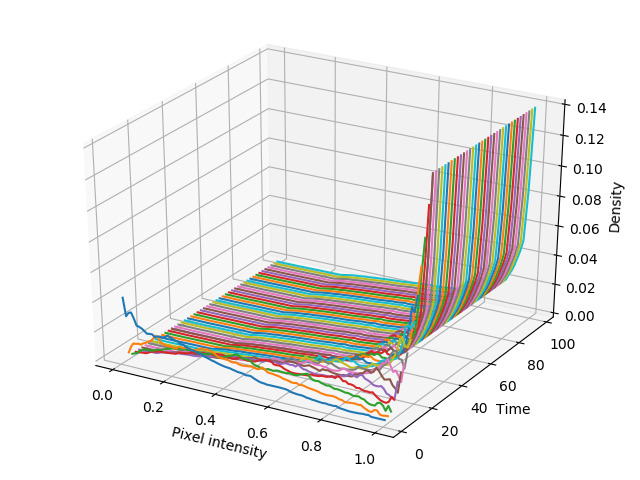}
    \caption{Comparison of the simulation results considering manually identified green areas and automatically identified. In the first column, the satellite images, in the second the diffusion results of the manual identification of green regions and in the third column, the results of the automatically identified green regions. Each row represents a different case.}
    \label{fig:5x5tiles}
\end{figure}

To validate the segmentation of the green areas, five figures of $5 \times 5$ tiles from the analyzed city, Ribeirão Preto, were manually annotated and the diffusion as described in the manuscript was simulated in the two cases, manually annotated and automatically identified green regions. Figure~\ref{fig:5x5tiles} shows the satellite image and the corresponding results of manually annotated and automatically identified green regions in the first and second columns respectively.

\begin{figure}[ht!]
    \centering
    \begin{subfigure}[b]{0.24\textwidth}
        \includegraphics[width=\textwidth]{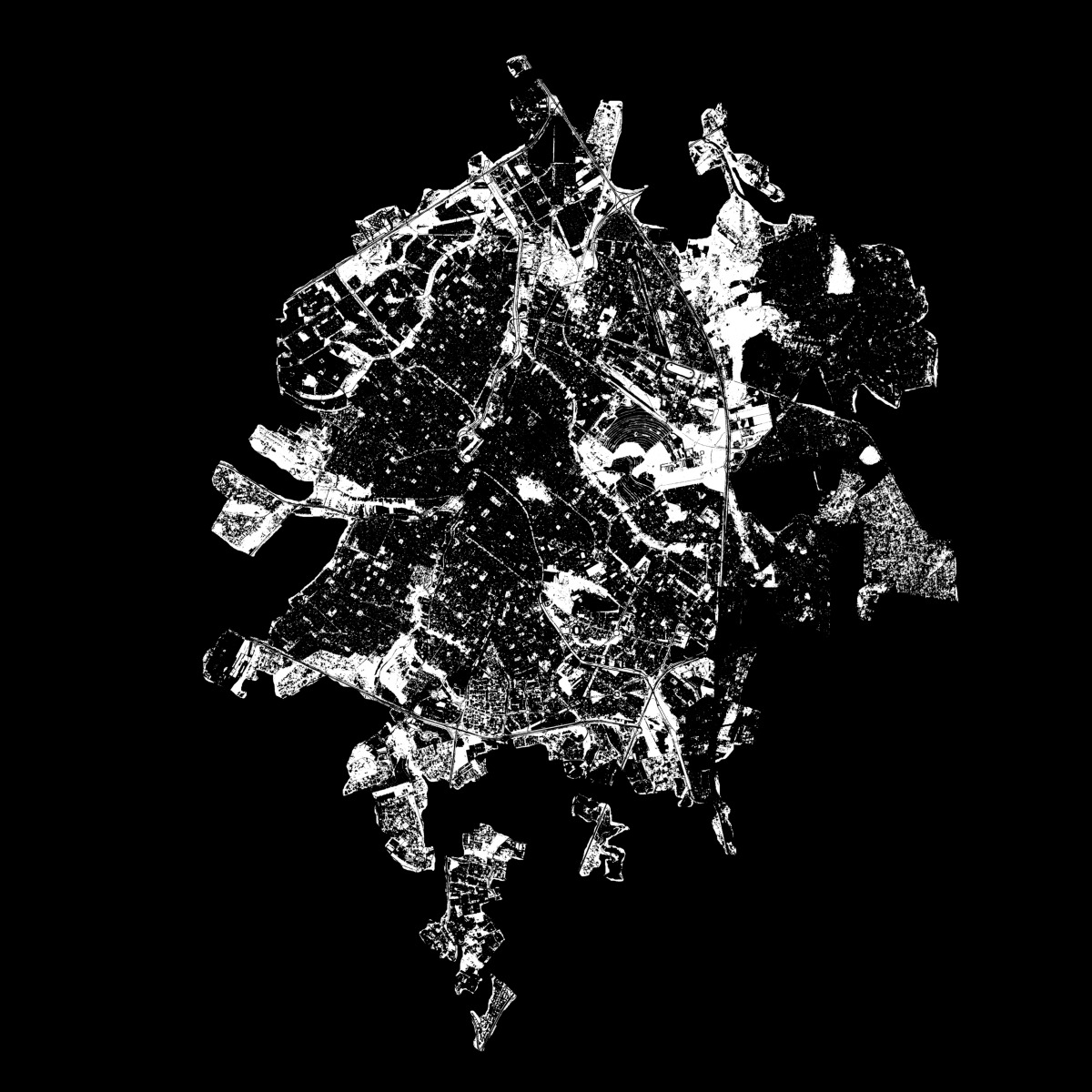}
    \end{subfigure}
    \begin{subfigure}[b]{0.24\textwidth}
        \includegraphics[width=\textwidth]{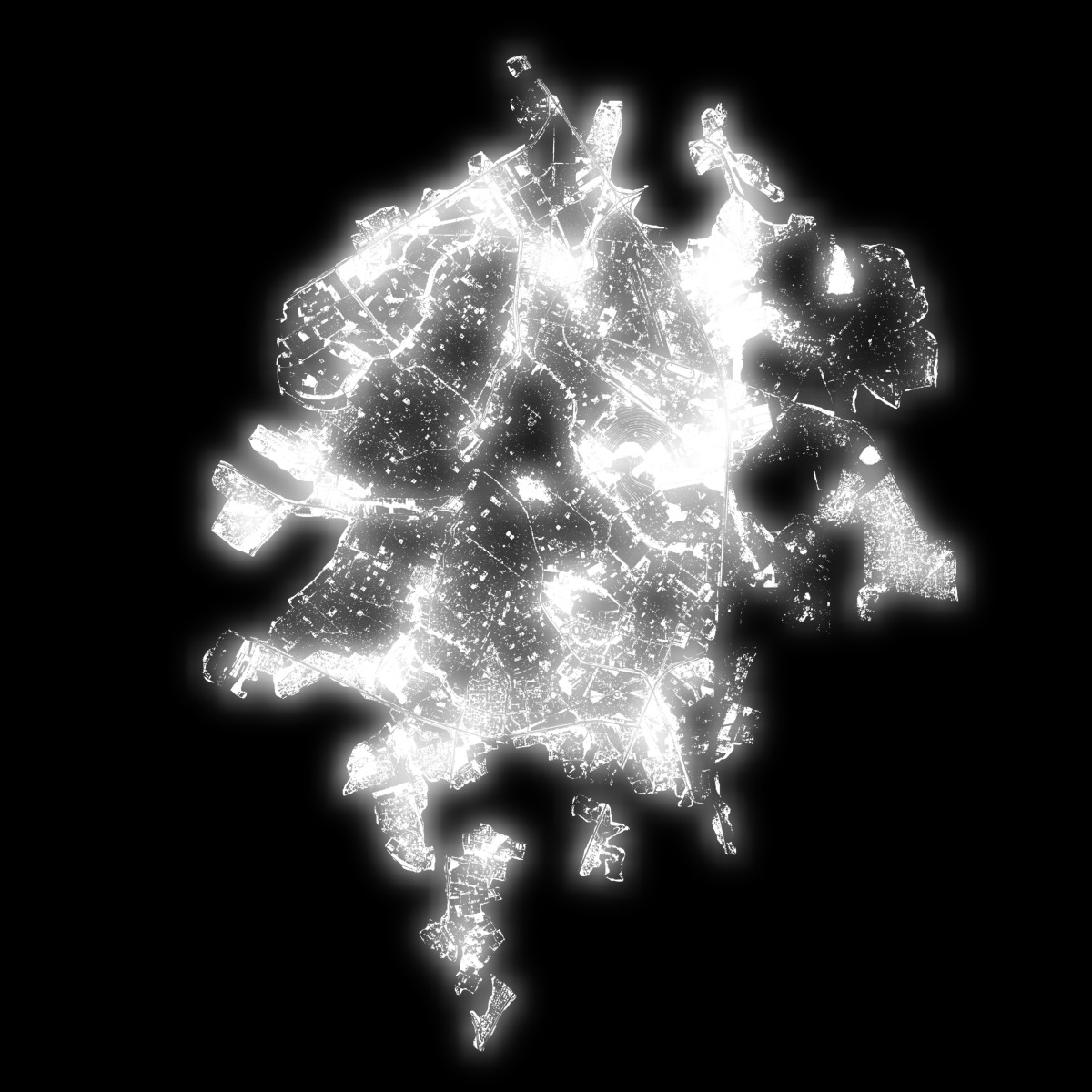}
    \end{subfigure}
    \begin{subfigure}[b]{0.24\textwidth}
        \includegraphics[width=\textwidth]{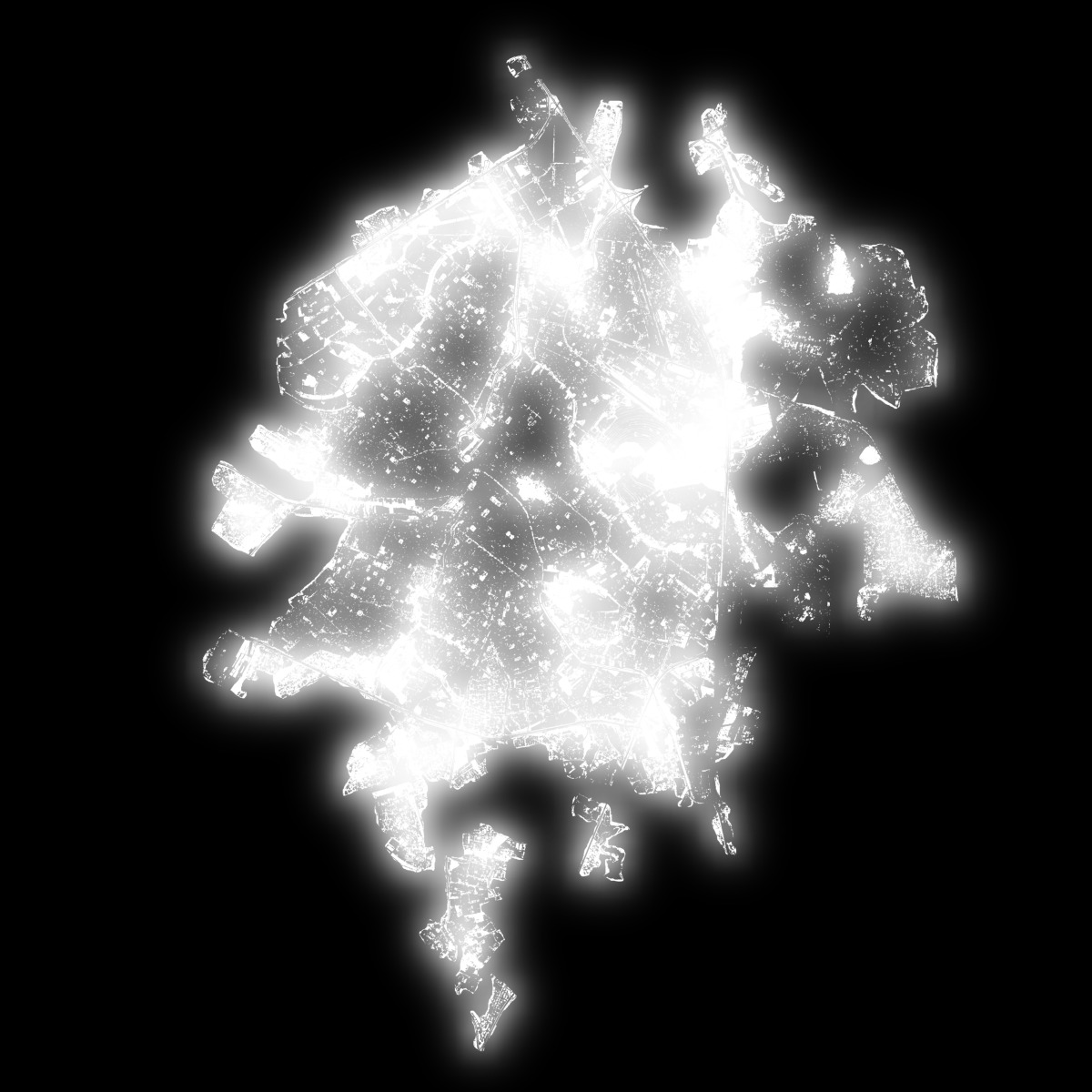}
    \end{subfigure}
    \begin{subfigure}[b]{0.24\textwidth}
        \includegraphics[width=\textwidth]{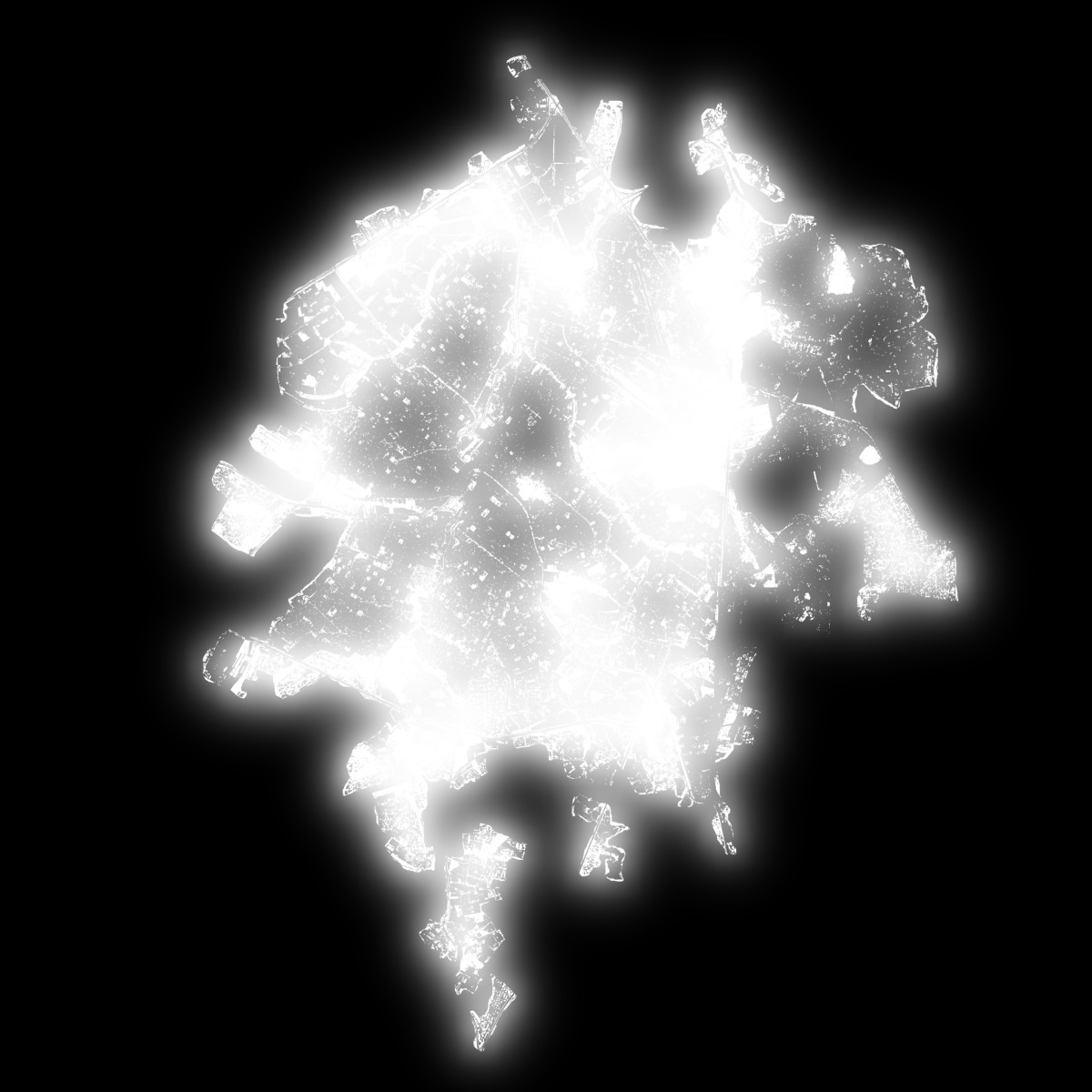}
    \end{subfigure}\\
    \begin{subfigure}[b]{0.24\textwidth}
        \includegraphics[width=\textwidth]{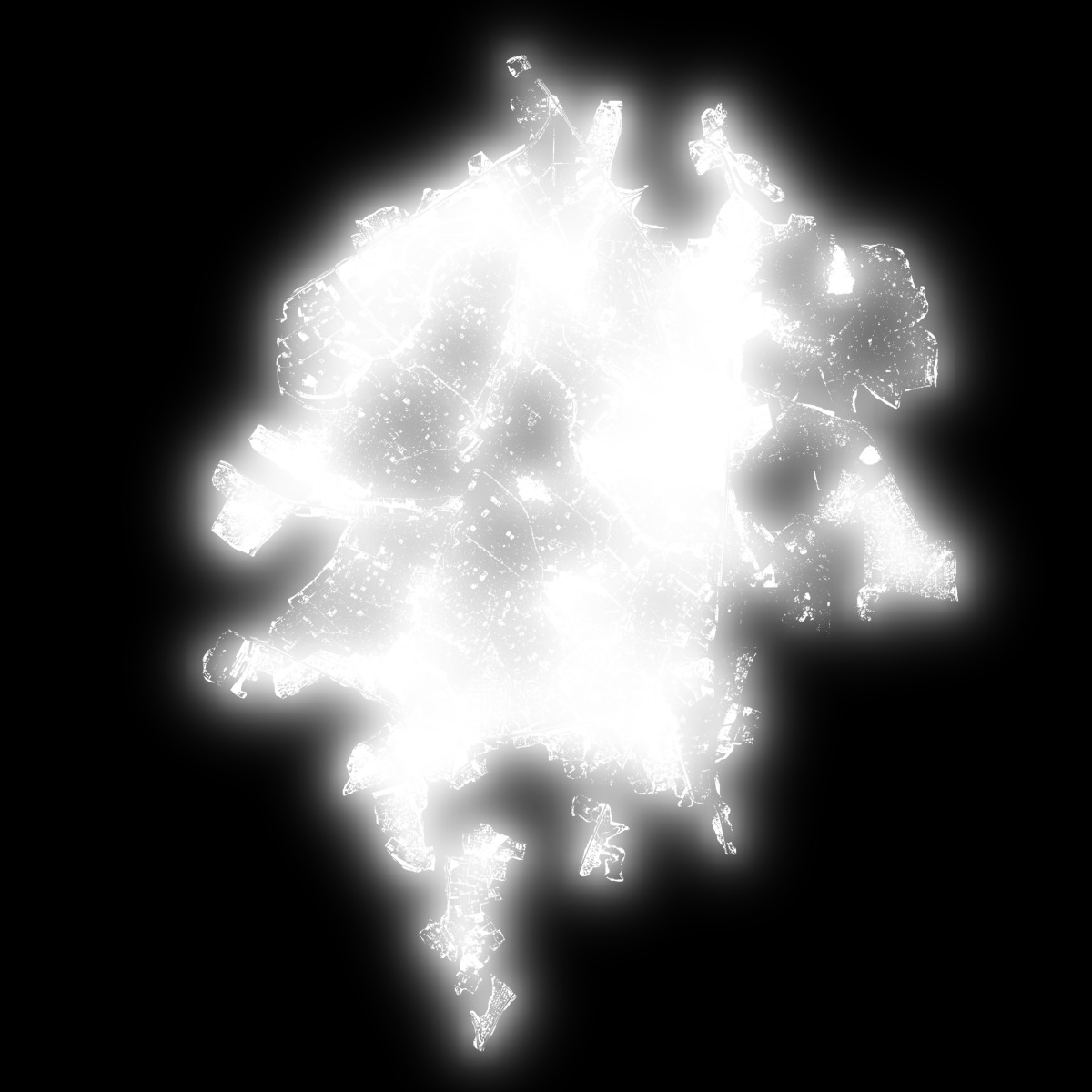}
    \end{subfigure}
    \begin{subfigure}[b]{0.24\textwidth}
        \includegraphics[width=\textwidth]{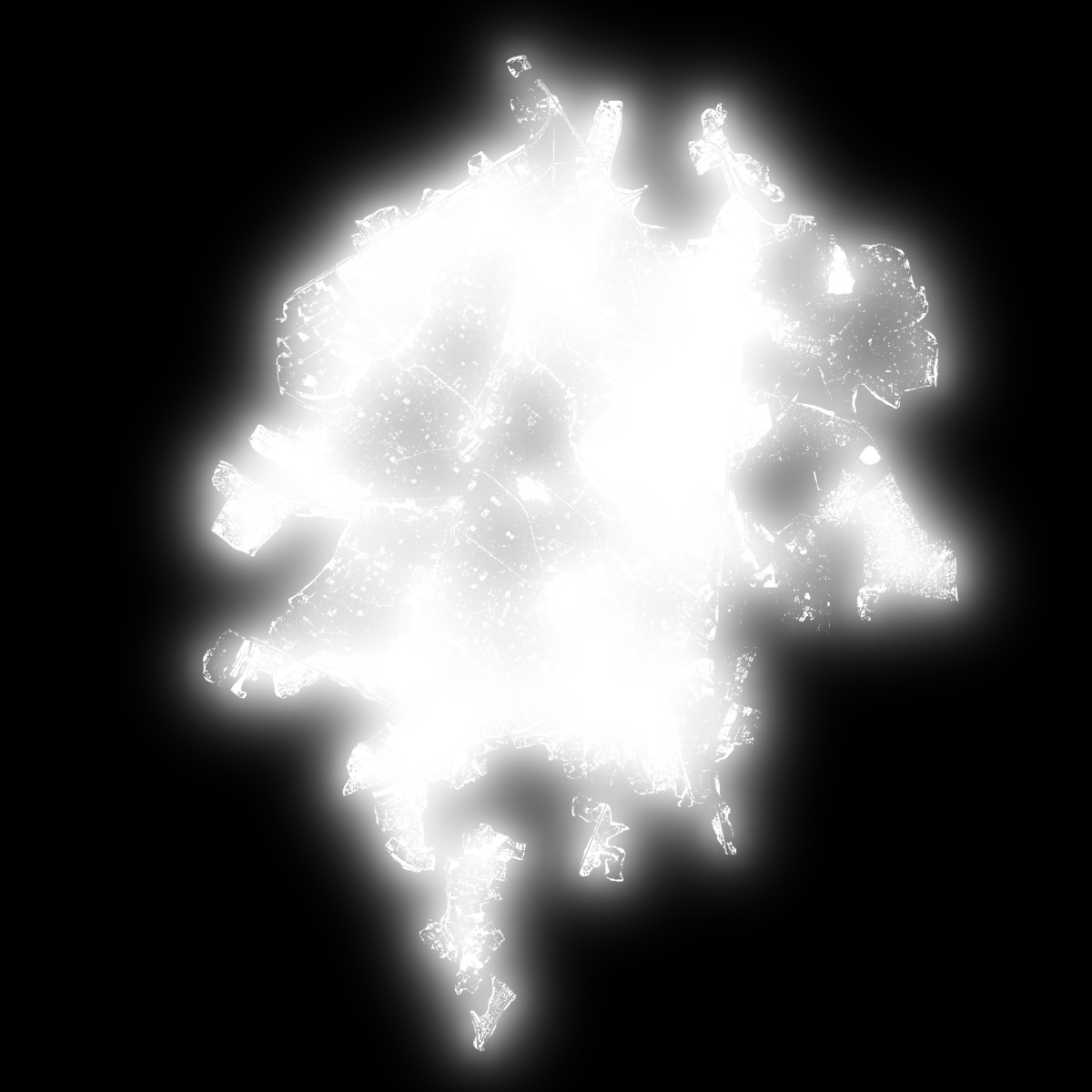}
    \end{subfigure}
    \begin{subfigure}[b]{0.24\textwidth}
        \includegraphics[width=\textwidth]{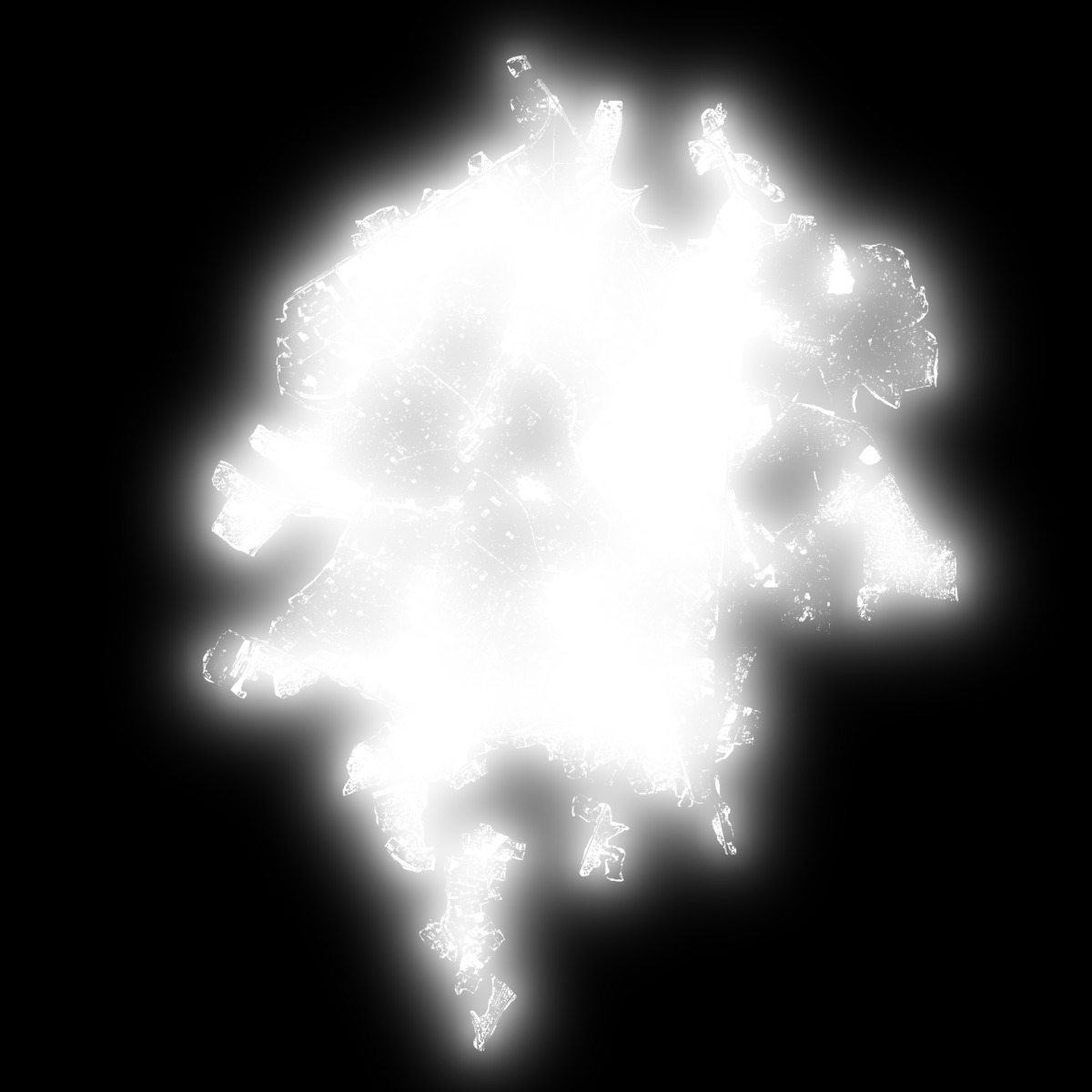}
    \end{subfigure}
    \begin{subfigure}[b]{0.24\textwidth}
        \includegraphics[width=\textwidth]{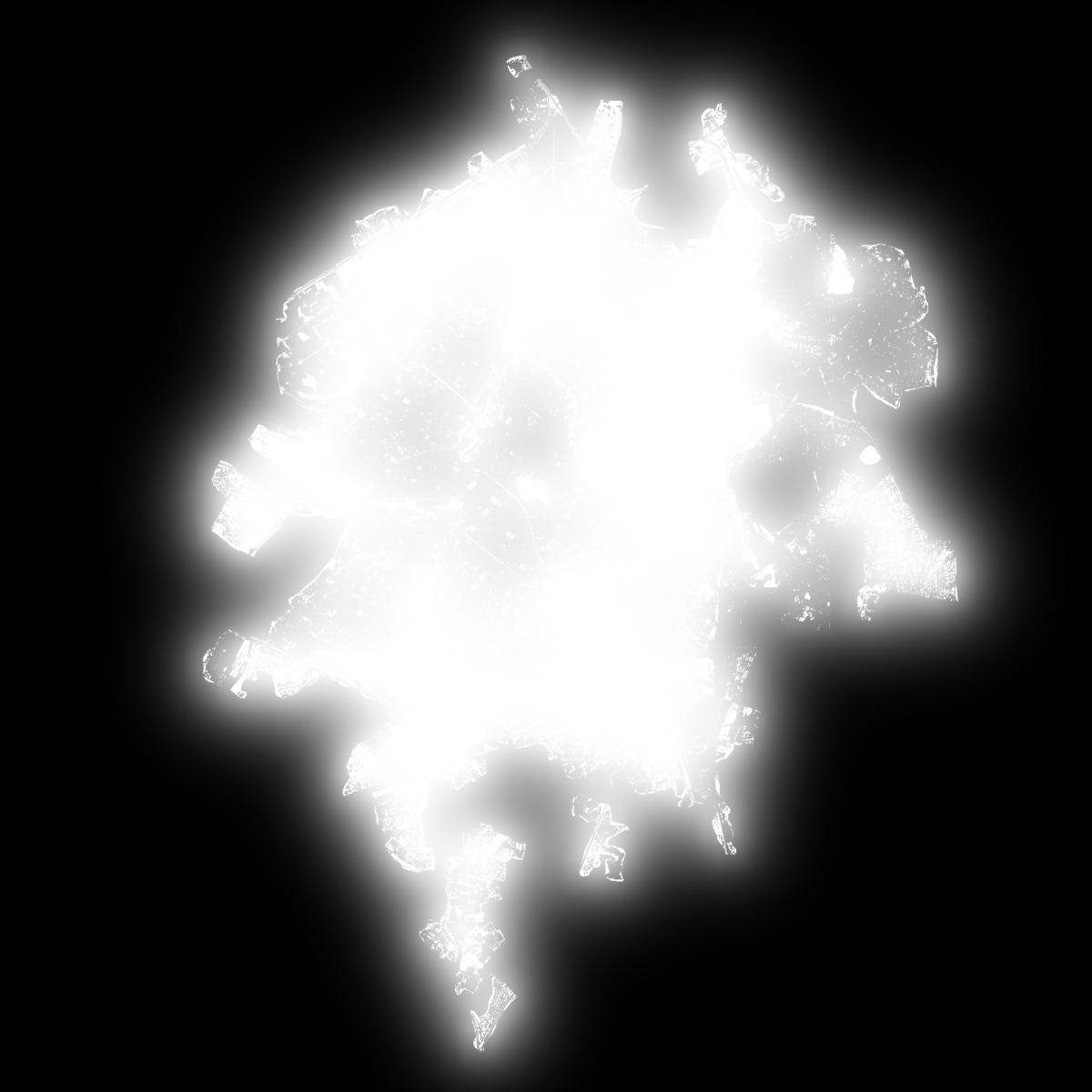}
    \end{subfigure}\\
    \caption{Visualization of the green diffusion in different steps of the green diffusion. It starts in the first row, from left to right, at instants 0, 3, 6 and 9 and continues in the second row, at instants 12, 15, 18 and 21.}
    \label{fig:diffusionsteps}
\end{figure}

Figure~\ref{fig:diffusionsteps} depicts the intermediate steps of the diffusion in steps 0, 3, 6, 9, 12, 15, 18, and 21.

